\documentclass[10pt]{iopart}
\usepackage{graphicx, iopams, harvard, algorithmicx, booktabs, varwidth}
\usepackage[noend]{algpseudocode}

\newcommand{\xvec}{\bi{x}}

\newcommand{\bvec}{\bi{b}}
\newcommand{\yvec}{\bi{y}}
\newcommand{\vvec}{\bi{v}}
\newcommand{\zvec}{\bi{z}}
\newcommand{\evec}{\bi{e}}
\newcommand{\muvec}{\boldsymbol \mu}
\newcommand{\norm}[1]{\left\Vert#1\right\Vert}
\begin{document}

\title[Superiorized algorithm for sparse-view and limited-angle polyenergetic CT]{Superiorized algorithm for reconstruction of CT images from sparse-view and limited-angle polyenergetic data}

\author{T.~Humphries$^1$, J. Winn$^1$, and A.~Faridani$^2$}

\address{$^1$ Division of Engineering and Mathematics, University of Washington Bothell, Bothell, WA 98011}

\address{$^2$  Department of Mathematics, Oregon State University, Corvallis,  OR 97331}

\ead{thumphri@uw.edu}

\begin{abstract} Recent work in CT image reconstruction has seen increasing interest in the use of total variation (TV) and related penalties to regularize problems involving reconstruction from undersampled or incomplete data. Superiorization is a recently proposed heuristic which provides an automatic procedure to ``superiorize'' an iterative image reconstruction algorithm with respect to a chosen objective function, such as TV. Under certain conditions, the superiorized algorithm is guaranteed to find a solution that is as satisfactory as any found by the original algorithm with respect to satisfying the constraints of the problem; this solution is also expected to be superior with respect to the chosen objective.

Most work on superiorization has used reconstruction algorithms which assume a linear measurement model, which in the case of CT corresponds to data generated from a monoenergetic X-ray beam. Many CT systems generate X-rays from a polyenergetic spectrum, however, in which the measured data represent an integral of object attenuation over all energies in the spectrum. This inconsistency with the linear model produces the well-known beam hardening artifacts, which impair analysis of CT images.

 In this work we superiorize an iterative algorithm for reconstruction from polyenergetic data, using both TV and an anisotropic TV (ATV) penalty. We apply the superiorized algorithm in numerical phantom experiments modeling both sparse-view and limited-angle scenarios. In our experiments, the superiorized algorithm successfully finds solutions which are as constraints-compatible as those found by the original algorithm, with significantly reduced TV and ATV values. The superiorized algorithm thus produces images with greatly reduced sparse-view and limited angle artifacts, which are also largely free of the beam hardening artifacts that would be present if a superiorized version of a monoenergetic algorithm were used.

%Recent work in CT image reconstruction has seen increasing interest in the use of total variation (TV) minimization to reconstruct images from sparse-view projection data. This interest is motivated primarily by the desire to reduce dose to the patient, as well as by recent results in the field of compressive sensing which provide guarantees of recoverability from undersampled data, under certain conditions. Following the vast majority of existing research in compressive sensing, these image reconstruction approaches typically assume a linear measurement model, which corresponds to data generated from a monoenergetic X-ray beam. Many clinical CT systems generate X-rays from a polyenergetic spectrum, however, which is inconsistent with a linear system model. This inconsistency produces the well-known beam hardening artifacts, which impair analysis of CT images.
%
%In this work we incorporate a polyenergetic, nonlinear projection model within the TV minimization framework, with the goal of producing images from sparse-view, polyenergetic data that are free of artifacts due to both undersampling and beam hardening. Our approach is based on two previously proposed iterative techniques for sparse-view and polyenergetic reconstruction, respectively. In numerical experiments we demonstrate that the proposed iterative method successfully eliminates both types of artifact, and that it outperforms a second approach combining analytical reconstruction with a classical beam hardening correction technique.
\end{abstract}

\noindent{\it Keywords}: computed tomography, superiorization, total variation, beam hardening

%\submitto{\PMB}
%\maketitle

\section{Introduction} 

Superiorization~\cite{CDH10,HGDC12} is an optimization heuristic in which an iterative algorithm for solving an inverse problem is {\em superiorized} by perturbing the solution within each iteration, in order to improve it with respect to some objective function. The goal of the superiorized algorithm is to produce a solution that satisfies the constraints of the inverse problem to the same extent as any solution produced by the original algorithm, which is also superior with respect to this objective function. The authors prove in~\cite{HGDC12} that this can be achieved provided that the original algorithm is {\em strongly perturbation resilient}, i.e., satisfies certain mathematical conditions that we describe later.

Concurrently, a great deal of recent work in computed tomography (CT) has focused on incorporating total variation (TV) minimization into iterative reconstruction. One application that has attracted a great deal of interest is sparse-view imaging, in which the image is reconstructed from fewer views than would be acquired in a typical scan. Sparse-view imaging is motivated primarily by the desire to reduce dose to the patient~\cite{BSHS10,MCKL12}, as well as by results from the field of compressive sensing, which state that signals that are sparse in some sense can be recovered from significantly fewer measurements than classical sampling theory dictates.  A major result in compressive sensing is that under certain conditions on the measurement model (e.g. that it satisfies a restricted isometry property, or RIP), finding the sparsest signal that satisfies the measured data is equivalent to the much more tractable problem of finding the signal with the smallest $\ell_1$ norm. Early papers developing these results include~\cite{CRT06a,CRT06b,D06b}, and comprehensive treatments of the topic can be found, for example, in~\cite{EK12,FR13}.

In CT imaging, it may be reasonable to assume that the image is piecewise constant with a relatively small number of edges, meaning that its discrete gradient is sparse. Total variation is simply the $\ell_1$ norm of the vector of discrete gradient magnitudes, and so TV minimization can therefore be used to find the image with the sparsest gradient which satisfies the measurements~\cite{NW13b}. It must be noted, however, that the system matrices encountered in CT imaging have not been proven to satisfy necessary conditions, such as the RIP, for these theoretical results to be applicable~\cite{JSP13,JS15}.

Superiorization does not claim to find a constraints-compatible solution that is {\em optimal} with respect to a chosen objective (e.g. sparsity), but rather a solution that is simply {\em superior} to the solution found by the original algorithm. The original algorithm is typically an iterative algorithm for solving large systems of linear equations, such as the algebraic reconstruction technique (ART) or one of its variants. The main appeal of superiorization is that it provides a straightforward method for modifying iterative algorithms in order to produce superior solutions, as well as theoretical guarantees that this is possible under certain circumstances. Total variation is a common choice of objective function, especially for studies on sparse-view CT; some examples of algorithms which have been applied in this context are ASD-POCS~\cite{SP08,BSHS10,HBRSP12,BYBH14}, and methods presented in~\cite{HD08,CDH10,HGDC12}. Other approaches for incorporating TV minimization into image reconstruction are also possible, and have been demonstrated in papers such as~\cite{TNC09,CWZS10,DVL11,RBFK11,SJP12}. 

Aside from TV, algorithms have been superiorized using other objectives such as the $\ell_1$ norm of the Haar transform~\cite{GHD11} and maximum entropy~\cite{DHC09}, though with less successful results. An anisotropic variant of TV was used to superiorize simultaneous ART (SART) in~\cite{CJLW13} for the case of limited-angle CT, in order to exploit the directional nature of artifacts that occur in this case. Superiorization has also been applied to problems in proton computed tomography~\cite{PSCR10}, positron emission tomography~\cite{GH14}, and radiation therapy~\cite{DCSG15}.

The goal of this paper is to present and validate a superiorized algorithm for reconstruction of CT images from polyenergetic data. Algorithms for CT image reconstruction which use a linear measurement model (such as ART) implicitly assume that measurements are generated from a monoenergetic X-ray beam. Many clinical CT systems generate X-rays from a polyenergetic spectrum, which produces measured data that is a nonlinear function of the object attenuation. The inconsistency with the linear model produces the well-known beam hardening artifacts, such as cupping and streaking~\cite{BD76}. These artifacts were first noted in early images of the brain~\cite{JS78}, and have also been shown to impede diagnosis in renal and cardiac imaging~\cite{BHLB07,RRDR10,KGALL10}. Many techniques for addressing beam hardening have been developed over the years, including corrections applied to the measured data or the reconstructed image~\cite{JS78,H79,HT83,JR97,KMPK10}, as well as iterative reconstruction algorithms which incorporate X-ray polychromaticity directly into the imaging model~\cite{HMDJ00,DNDMS01,EF02,VVDB11,LS14b}.

The algorithm we choose to superiorize is polyenergetic SART (pSART)~\cite{LS14b}, which uses interpolation to model the attenuation map of the object at all energies as a function of the attenuation at a reference energy. We apply the superiorized algorithm in numerical phantom experiments modeling sparse-view and limited-angle reconstruction scenarios, using both TV and the anisotropic variant described in~\cite{CJLW13}. While we are not able to prove that pSART is perturbation resilient -- and hence that a superiorized version is guaranteed to produce solutions that are as constraint-compatible as the original algorithm -- our superiorized version of pSART does produce solutions of this type in these numerical experiments. 

%The paper is organized as follows. In Section~\ref{S:methods} we describe the algorithms and mathematical models used in the paper. Section~\ref{S:results} gives results of numerical phantom simulations which show that the superiorized algorithm produces images from sparse-view and limited-angle polyenergetic data which are superior with respect to the chosen objective, and have significantly reduced artifacts as a result. The main findings of the paper are summarized in Section~\ref{S:conclusions}.

\section{Methodology}\label{S:methods}
We first describe the general framework for superiorization of an iterative method. This is followed by a discussion of the mathematical model used for both monoenergetic and polyenergetic CT data, a description of the pSART method, and finally our description of the superiorized version of pSART.

\subsection{Superiorization}

The framework for superiorization is described in detail in~\cite{HGDC12}; here we summarize the main points from that paper. As an illustrative example, we consider the problem of finding a non-negative solution to a linear system of equations $A \xvec = \bvec$ with $I$ equations for $J$ unknowns. The domain of the problem is the set of all vectors with $J$ non-negative components, $\Omega = R^J_+$. A {\em problem set}, $\mathbb{T}$, is the set of all possible matrices $A \in \mathbb{R}^{I \times J}$ and right-hand sides $\bvec \in \mathbb{R}^I$ which describe an inverse problem of this type; a {\em problem}, $T \in \mathbb{T}$, refers to the constraints imposed by a specific matrix and right-hand side. 

Since it is often impossible to satisfy all constraints of a problem exactly, we define a {\em proximity function}, $Pr$, which assigns to every problem $T$ an operator $Pr_T: \Omega \to \mathbb{R}_+$ that quantifies the extent to which a solution satisfies the constraints of the problem. A simple choice is the residual:
\begin{equation}
Pr_T(\xvec) = \norm{\bvec - A \xvec}_2. \label{E:res_mono}
\end{equation}
We then define an {\em $\varepsilon$-compatible} solution to the problem as any solution such that $Pr_T(\xvec) \leq \varepsilon$, where $\varepsilon \geq 0$.

Together, $\mathbb{T}$ and $Pr$ form a {\em problem structure}. An {\em algorithm}, $\mathbf{R}$, for a given problem structure assigns an operator $\mathbf{R}_T: \Delta \to \Omega$ to every problem $T$. The set $\Delta$ is such that $\Omega \subseteq \Delta \subseteq \mathbb{R}^J$; for example, $\mathbf{R}_T$ could take as input vectors in $\mathbb{R}^J$ with negative components, even if the desired output has only non-negative components. From any initial point $\xvec^{(0)}$, this operator produces a sequence of iterates,
\begin{equation*}
\left( (\mathbf{R}_T)^k \xvec^{(0)} \right)_{k=0}^\infty = \xvec^{(0)}, \mathbf{R}_T ( \xvec^{(0)} ), \mathbf{R}_T (\mathbf{R}_T (\xvec^{(0)})), \dots. 
\end{equation*}  

The algorithm $\mathbf{R}$ is said to be {\em strongly perturbation resilient} if the following two conditions are satisfied for every problem $T \in \mathbb{T}$:
\begin{enumerate}
\item There exists an $\varepsilon > 0$ such that $\mathbf{R}_T$ eventually produces an $\varepsilon$-compatible solution, from all starting points $\xvec^{(0)} \in \Omega$, and
\item For all $\varepsilon \in \mathbb{R}_+$ for which $\mathbf{R}_T$ is able to produce an $\varepsilon$-compatible solution, the modified iteration
\begin{equation}
\xvec^{(k+1)} = \mathbf{R}_T \left(\xvec^{(k)} + \beta_k \vvec^{(k)} \right) \label{E:perturb}
\end{equation} 
produces a solution that is $\varepsilon'$-compatible, for any $\varepsilon' > \varepsilon$. Here $(\beta_k)^\infty_{k=0}$ represents a summable sequence of non-negative real numbers, and $(\vvec^{(k)})^\infty_{k=0}$ is a bounded sequence of vectors $\vvec^{(k)} \in \mathbb{R}^J$.
\end{enumerate}
Essentially these conditions state that even if the solutions generated by $\mathbf{R}_T$ are perturbed after every iteration (subject to some restrictions on the size of these perturbations), one is able to obtain a solution that is equally as constraints-compatible as that obtained by the unperturbed algorithm. %The fact that these perturbations could cause the solution to leave the problem domain $\Omega$ is the reason why it is desirable for the domain of $\mathbf{R}_T$ to be the superset $\Delta$.

The authors of~\cite{HGDC12} give some sufficient conditions on $\mathbf{R}$ and $Pr$ for the algorithm to be strongly perturbation resilient. The essential idea of superiorization is that the perturbation directions $\vvec^{(k)}$ can be chosen as nonascending directions of an objective function, $\phi : \Delta \to \mathbb{R}$. Thus, any strongly perturbation resilient algorithm can be ``steered'' according to this objective without jeopardizing its convergence towards an $\varepsilon$-compatible solution. We discuss how this is achieved practically in Section~\ref{S:pSARTsup}.%In Section~\ref{S:pSARTsup} we show how this is implemented in practice for pSART, but first we describe the mathematical model for CT imaging used in this paper.

\subsection{Mathematical model}
For the sake of simplicity we restrict our discussion to the two-dimensional case. Let $\mu(\yvec, E): \mathbb{R}^2 \times \mathbb{R} \to \mathbb{R}$ denote the object attenuation as a function of position $\yvec$ and energy $E$. For a monoenergetic X-ray beam with energy $E_0$, the measurement along a line $\ell$ is given by
\begin{eqnarray}
I_\ell = I_0 \exp \left( -\int_\ell \mu(\yvec, E_0) \: d \yvec \right), \textrm{ or} \label{E:mono1} \\
-\ln \left( I_\ell / I_0 \right) =  \int_\ell \mu(\yvec, E_0) \: d \yvec \label{E:mono2}
\end{eqnarray}
where $I_\ell$ is the measured intensity and $I_0$ is the initial intensity of the beam. Suppose that $\mu(\yvec, E_0)$ is discretized as an $n \times n$ pixel image, and we collect a total of $I$ measurements. Letting $J=n^2$, the data can be represented as a linear system of equations
\begin{equation}
\bvec = A \xvec, \label{E:mono3}
\end{equation}
where $\xvec \in \mathbb{R}^{J}_+$ is a vector representing the image of $\mu(\yvec, E_0)$, $\bvec \in \mathbb{R}^I_+$ is the log-transformed projection data (left side of \eref{E:mono2}), and $A \in \mathbb{R}^{I \times J}$ is the system matrix whose $(i,j)$th element is the length of intersection of the $i$th line with the $j$th pixel of $\xvec$. 

If the X-ray beam is polyenergetic, \eref{E:mono1} becomes
\begin{equation}
I_\ell = \int S(E)   \exp \left( -\int_\ell \mu(\yvec, E) \: d \yvec \right) d E,
\end{equation}
where $S(E)$ represents the spectrum of the beam as a function of energy. For the purpose of image reconstruction, both the line integral and the integral with respect to energy are discretized, giving
\begin{equation}
I_\ell = \sum_{h=1}^{N_h} S_h \exp \left( -a_i \xvec_h \right),\label{E:poly}
\end{equation}
where $a_i$ is the row of $A$ corresponding to measuring along line $\ell$, $N_h$ is the number of quadrature points (i.e., energies) used in the discretization, $S_h$ is the quadrature weight corresponding to energy $E_h$, and $\xvec_h$ is the discretized image of $\mu(\yvec, E_h)$. Without additional modeling, the image reconstruction problem is not tractable as it requires reconstructing the images $\xvec_h$ at $N_h$ different energy levels, from a single set of tomographic data. Thus, techniques for reconstruction of images from polyenergetic data must typically include a model for how $\mu(\yvec, E)$ varies with energy. We discuss the model used by pSART in the next section.

\subsection{Polyenergetic SART}
We first describe the simultaneous algebraic reconstruction technique (SART)~\cite{AK84}, an iterative algorithm for image reconstruction based on the linear model~\eref{E:mono3}. Following~\cite{CE02}, we define diagonal matrices $D \in \mathbb{R}^{J \times J}$ and $M \in \mathbb{R}^{I \times I}$ with
\begin{eqnarray}
D_{jj} &= \frac{1}{\zeta_j}, ~~\zeta_j = \sum_{k=1}^m |a_{kj}|, j = 1 \dots J \\
M_{ii} &= \frac{1}{\eta_i}, ~~\eta_i = \sum_{k=1}^n |a_{ik}|, i = 1 \dots I. \label{E:diagmatrices} 
\end{eqnarray}
In other words, $\zeta_j$ is the 1-norm of the $j$th column of $A$, and $\eta_i$ is the 1-norm of the $i$th row. Furthermore, we organize the measurements corresponding to each angular view into $N_w$ subsets, denoted by $s(w)$ for $w = 1 \dots N_w$. Let $A_{s(w)}$ denote the matrix obtained by extracting only the rows of $A$ corresponding to $s(w)$, and similarly for $D_{s(w)}$, $M_{s(w)}$ and the vector of measurements, $\bvec_{s(w)}$. Then, a block-iterative algorithm for solving the problem $T$ described by~\eref{E:mono3} is given by
\begin{eqnarray}
\mathbf{R}_T(\xvec) = \mathbf{Q} \mathbf{B}_{N_w} \dots \mathbf{B}_2 \mathbf{B}_1 (\xvec), \label{E:SART1}
\end{eqnarray}
where
\begin{eqnarray}
\mathbf{B}_w (\xvec) = \xvec  - D_{s(w)} (A_{s(w)})^T M_{s(w)}\left[ A_{s(w)} \xvec - \bvec_{s(w)} \right],\label{E:SART2}
\end{eqnarray}
and
\begin{eqnarray}
(\mathbf{Q} \xvec)_j = \max\{0, x_j\},~~ j \in [1, J]. \label{E:SART3}
\end{eqnarray}
Here $\mathbf{B}_w$ is an iterative step aiming to satisfy the constraints imposed by the $w$th subset of measurements, and $\mathbf{Q}$ is an operator that sets negative values of $\xvec$ to zero. The subsets can be chosen in several ways. For a parallel-beam geometry consisting of $N_v$ angular views of the object, the case where $N_w = N_v$ (i.e., one view per subset) corresponds to the classical version of SART~\cite{AK84}, while the case where $N_w=1$ (all views are processed simultaneously) is typically referred to as SIRT (simultaneous iterative reconstruction technique). In this paper we use an ordered subsets approach~\cite{HL94}, where each subset consists of $N_v / N_w$ equally spaced views around the object.

Polyenergetic SART (pSART)~\cite{LS14b} replaces the monoenergetic forward model used by SART with a log-transformed polyenergetic forward projection operator $\mathcal{P}: \mathbb{R}_+^{J} \to \mathbb{R}^I_+$, given by:
\begin{equation}
[\mathcal{P} (\xvec)]_i = - \ln \left[ \sum_{h=1}^{N_h} S_h \exp \left( - a_i \muvec (\xvec, E_h)  \right) \right] \biggl/ \sum_{h=1}^{N_h} S_h. \label{E:project_poly}
\end{equation}
The image to be reconstructed, $\xvec$, is the attenuation map at some reference energy $E_0$. The function $\muvec: \mathbb{R}_+^J \times \mathbb{R} \to \mathbb{R}_+^J$ maps $\xvec$ to $\xvec_h$ (cf. \Eref{E:poly}) using linear interpolation between tabulated attenuation curves for basis materials such as air, fat, soft tissue and bone. In particular, if $x_j$ denotes the linear attenuation coefficient (LAC) of pixel $j$ at $E_0$, then the attenuation coefficient of that pixel at all other energies is computed as
\begin{equation}
\mu(x_j, E) = \frac{ [\mu_{m+1}(E_0) - x_j] \mu_{m}(E) +  [x_j - \mu_{m}(E_0) ] \mu_{m+1}(E)}{\mu_{m+1}(E_0) - \mu_{m}(E_0)}, \label{E:mu_interp}
\end{equation}
where $\mu_m(E)$ and $\mu_{m+1}(E)$ are the energy-dependent LAC functions for the two basis materials with LAC values adjacent to $x_j$ at the reference energy. So, if a pixel's intensity is halfway betwen the LAC for fat and soft tissue at the reference energy, for example, then it is assumed that its value is halfway between the LAC for fat and soft tissue at all other energies as well.

The pSART algorithm can be expressed in the same way as algorithm~\eref{E:SART1}, but $\mathbf{B}_w$ changed to
\begin{eqnarray}
\mathbf{B}_w (\xvec) = \xvec  - D_{s(w)} (A_{s(w)})^T M_{s(w)}\left[ \mathcal{P}_{s(w)}( \xvec) - \bvec_{s(w)} \right],\label{E:pSART1}
\end{eqnarray}
where $\mathcal{P}_{s(w)}$ is the same as~\eref{E:project_poly} but using only rows of $A_{s(w)}$ rather than $A$, and the elements of $\bvec$ are given by
\begin{equation}
b_\ell = -\ln \left( I_\ell \biggl/ \sum_{h=1}^{N_h} S_h \right).\label{E:pSART2}
\end{equation}
Finally, the proximity function is the residual of the log-transformed data (cf.~\eref{E:res_mono}):
\begin{equation}
Pr_T(\xvec) = \norm{\bvec - \mathcal{P}(\xvec)}_2, \label{E:res_poly}
\end{equation} 

The pSART algorithm is a nonlinear fixed point iteration, whose convergence properties are not well-established. Our previous work in~\cite{H15} shows that, in constrast to SART, convergence of the algorithm cannot be established in general. Numerical and physical phantom experiments in~\cite{LS14b,H15}, however, indicate that it works well in practice, i.e., with realistic system matrices and measurements. We have chosen to superiorize pSART in this paper due to its similarity to SART and other algorithms that have been successfully superiorized.

\subsection{Superiorized pSART}\label{S:pSARTsup}

%Using our previous notation, the problem $T$ for polyenergetic reconstruction using pSART consists of the system matrix $A$, the postlog measurements $\bvec$~\eref{E:pSART2}, the discrete approximation to the spectrum, $S_h$, and the tabulated attenuation curves used in \eref{E:mu_interp}. The proximity function is the post-log residual
%\begin{equation}
%Res_T(\xvec) = \norm{\bvec - \mathcal{P}(\xvec)}_2,
%\end{equation}
%and the algorithm is the iteration~\eref{E:SART1} with $\mathbf{B}_w$ given by~\eref{E:pSART1}.

We consider superiorization with respect to two objective functions, total variation (TV) and anisotropic TV (ATV). Suppose that $\xvec$ is represented as a two-dimensional image, with $x_{m,n}$ denoting the pixel in the $m$th row and $n$th column of $\xvec$. We define a discrete, local gradient operator by
\begin{equation}
\nabla \xvec_{m,n} = \left[ x_{m+1,n}-x_{m,n},~x_{m,n+1}-x_{m,n} \right]^T,
\end{equation}
and can then define the TV as
\begin{eqnarray}
\phi_{TV}(\xvec) &= \sum_{m,n} \norm{ \nabla \xvec_{m,n} }_2 \nonumber \\
&= \sum_{m,n} \sqrt{ (x_{m+1,n}-x_{m,n})^2 + (x_{m,n+1}-x_{m,n})^2 + \epsilon^2},\label{E:TV1}
\end{eqnarray}
where $\epsilon$ is a small parameter which is added to ensure that the function is differentiable everywhere.  The partial derivatives of $\phi_{TV}$ with respect to $\xvec_{m,n}$ can be computed straightforwardly, and form the gradient vector $\nabla \phi_{TV} (\xvec) \in \mathbb{R}^J$.

%The partial derivatives of this function are given by
%\begin{eqnarray}
%\fl \frac{\partial}{\partial x_{m,n}} \phi_{TV} (\xvec) = &\frac{x_{m,n}-x_{m-1,n}}{\sqrt{ (x_{m,n}-x_{m-1,n})^2 + (x_{m-1,n+1}-x_{m-1,n})^2 + \epsilon^2}} + \nonumber \\ 
% 					&~~\frac{x_{m,n}-x_{m,n-1}}{\sqrt{ (x_{m+1,n-1}-x_{m,n-1})^2 + (x_{m,n}-x_{m,n-1})^2 + \epsilon^2}} + \nonumber \\ 					
% 					& ~~\frac{2 x_{m,n} - x_{m,n+1}-x_{m+1,n}}{\sqrt{ (x_{m+1,n}-x_{m,n})^2 + (x_{m,n+1}-x_{m,n})^2 + \epsilon^2}} \label{E:TV2},
%\end{eqnarray} and so the gradient vector $\nabla \phi_{TV} (\xvec) \in \mathbb{R}^J$ is the vector consisting of these partial derivatives, ordered in the same way as the components of $\xvec$ when it is represented as a vector.

The anisotropic TV (ATV) variant considered here was proposed in \cite{CJLW13} for limited-angle reconstruction. It is defined as
\begin{equation}
\phi_{ATV}(\xvec) = \sum_{i=1}^{N_\alpha} \omega_i \sum_{m,n} \norm{ \nabla_{\alpha_i} \xvec_{m,n}}_2, \label{E:ATV1}
\end{equation}
where
\begin{equation}
\nabla_{\alpha} \xvec_{m,n} = \left( \nabla \xvec_{m,n} \cdot \evec_{\alpha} \right) \evec_\alpha \label{E:ATV2}
\end{equation}
and $\evec_\alpha = [\cos \alpha,~\sin \alpha]^T$ is a unit vector which has angle $\alpha$ with the positive $x$-axis. The ATV is computed by selecting $N_\alpha$ directions $\evec_\alpha$ and projecting the discrete, local gradient at every point onto the corresponding unit vector. The sum of the 2-norms of those vectors gives the total variation in the direction of $\evec_\alpha$, and the total ATV is a weighted sum of these directional total variations. The weights, $\omega_i$, should sum to one and can be chosen to give higher weight along certain directions. In particular, reconstruction from limited-angle data produces artifacts whose directions are determined by the missing data (see Section~\ref{S:SVLA}), and so the $\omega_i$ can be chosen to have stronger weighting orthogonal to those directions. A heuristic for determining $\omega_i$ based on the missing angles, which we use in this paper, is described in~\cite{CJLW13}. As in \eref{E:TV1}, we include a small parameter $\epsilon$ when computing $\norm{ \nabla_{\alpha_i} \xvec_{m,n}}_2$ to ensure differentiability, and form $\nabla \phi_{ATV} (\xvec)$ from the partial derivatives with respect to $\xvec_{m,n}$.

%Combining \eref{E:ATV1} and \eref{E:ATV2} gives the following expression for $\phi_{ATV}$ and its gradient:
%{\bf insert expression here}

The superiorized version of pSART is presented in \Fref{F:pSART_sup}.  The inputs to the algorithm are: a starting point $\xvec^{(0)}$, parameter $N$ specifying the number of inner loop iterations to perform, parameter $\gamma \in (0, 1)$ controlling the step size for minimizing $\phi$, and target constraint-compatibility value $\varepsilon_{target}$ indicating when to terminate the algorithm. The objective function $\phi$ is either $\phi_{TV}$~\eref{E:TV1} or $\phi_{ATV}$ \eref{E:ATV1} in our experiments. The operators $\mathbf{B}_w$ are as specified in \eref{E:pSART1} and require as additional input the system matrix $A$, log-transformed measurements $\bvec$, number of projection subsets $N_w$, and the quadrature weights $S_h$, reference energy $E_0$, and basis material attenuation curves used in \eref{E:project_poly} and \eref{E:mu_interp}. $\mathbf{Q}$ is the non-negativity operator~\eref{E:SART3}.

\begin{figure}
\hrulefill
\begin{algorithmic}[1]
\State $\ell \gets -1$
\For{$k = 0, 1, 2, \dots$}
\State $\xvec^{(k,0)} \gets \xvec^{(k)}$ 
\For{$n = 0, 1, 2,  \dots N-1$}
\State $\vvec^{(k,n)} \gets -\nabla \phi (\xvec^{(k,n)}) / \left( \norm{ \nabla \phi(\xvec^{(k,n)}} + \delta \right)$
\While{true}
\State $\ell \gets \ell+1$
\State $\beta_{k,n} \gets \gamma^\ell$
\State $\zvec \gets \xvec^{(k,n)} + \beta_{k,n} \vvec^{(k,n)}$
\If {$\phi(\zvec) \leq \phi(\xvec^{(k)})$}
\State $\xvec^{(k,n+1)} \gets \zvec$
\State {\bf break}
\EndIf
\EndWhile
\EndFor
\State $\xvec^{(k+1)} \gets   \mathbf{Q} \mathbf{B}_{N_w} \dots \mathbf{B}_2 \mathbf{B}_1 (\xvec^{(k,N)})$ 
\If{$Pr_T(\xvec^{(k+1)}) < \varepsilon_{target}$}
\State \Return $\xvec^{(k+1)}$
\EndIf
\EndFor
%\State\Return $\xvec^{(k)}$
\end{algorithmic}
\hrulefill
\caption{Pseudocode of the superiorized pSART algorithm.}\label{F:pSART_sup}
\end{figure}

The inner loop over $n$ (lines 4 to 12) is the key component that is added to pSART to superiorize it. On line 5 we set  $\vvec^{(k,n)}$ to be the negative, normalized gradient at the point $\xvec^{(k,n)}$. The small parameter $\delta$ is included to ensure no division by zero. Lines 6 to 12 form another loop in which $\xvec^{(k,n)}$ is perturbed in the direction of $\vvec^{(k,n)}$ by a decreasing step size until a point $\zvec$ is found such that $\phi(\zvec) \leq \phi(\xvec^{(k)})$. One can prove inductively that since $\vvec^{(k,n)}$ is a nonascending vector of $\phi$ at $\xvec^{(k,n)}$, this loop must eventually terminate~\cite{HGDC12}. This point $\zvec$ is then used in the next iteration of the inner loop over $n$. Following exactly $N$ repetitions of this loop, the resulting point is passed as input to pSART (line 13). We then check to see whether the solution is $\varepsilon$-compatible for some pre-specified value $\varepsilon_{target}$, and return the solution if so. %(e.g. in the case where we initialize $\xvec^{(0)}$ as a constant vector, in which case $\nabla \phi$ is identically zero)

The stopping criterion for the algorithm is based on producing an $\varepsilon_{target}$-compatible solution, so one must ensure that it is possible to achieve such a solution if the algorithm is ever to terminate. The value of $\varepsilon_{target}$ can be obtained by taking the value of $\varepsilon$ corresponding to a solution produced by the unsuperiorized version of the algorithm; if the algorithm is strongly perturbation resilient, then the superiorized version will be able to find an equally compatible solution~\cite{HGDC12}. It is not clear, however, that pSART is actually a strongly perturbation resilient algorithm. Some sufficient conditions for strong perturbation resilience are given in~\cite{HGDC12}, which include that the algorithm is convergent for all problems; we have shown previously that this is not the case for pSART~\cite{H15}. In the numerical experiments of the next section, however, our superiorized version of pSART is able to produce solutions that are equally constraints-compatible to those found by pSART.

The superiorized pSART algorithm is similar to one previously presented by the authors in~\cite{HF15}, which we denoted as pSART-iTV. This algorithm also combined pSART with TV minimization, using a technique based on the improved TV (iTV) method of~\cite{RBFK11}. The technique used by iTV is similar to superiorization, but the size of the descent steps during the TV minimization process is controlled using a different heuristic. In this paper we have chosen to use a superiorized version of pSART following the technique of~\cite{HGDC12}. We will refer to the superiorized algorithm as pSART-TV or pSART-ATV, depending on the choice of the objective function $\phi$. %, and so it is not apparent that it is entirely equivalent to superiorization

We note that algorithms combining polyenergetic modeling and minimization of secondary objectives (such as TV) have been considered previously in some studies. The polyenergetic iterative method of~\cite{EF02} includes an edge-preserving Huber penalty which is similar to TV, though the primary purpose of this penalty was to denoise the image, and the paper did not consider sparse-view or limited-angle data. In \cite{BB15}, the authors use a compressive sensing approach to reconstruct sparse-view polyenergetic data, but used constraints based on wavelet sparsity and matrix factorization rather than TV. Finally, an algorithm for material decomposition from spectral CT data presented in~\cite{BSSP16} incorporates polyenergetic modeling within each energy window as well as a TV minimization constraint. In this paper we do not consider the case of spectral CT, in which photon-counting detectors allow for simultaneous acquisition of several data sets in different energy windows, but rather the more traditional model where the data represent an integral over the entire range of X-ray energies. Beam-hardening artifacts are likely to be more severe in this case as the attenuation coefficients of materials vary more over a larger energy range.

\section{Numerical Experiments}\label{S:results}
In this section we demonstrate the effectiveness of the superiorized pSART algorithm using numerical phantom experiments. As our numerical experiments are concerned with sparse-view and limited-angle data, we first define what is meant by these two terms. 
\subsection{Sparse-view and limited-angle data}\label{S:SVLA}
Recall that the set of all line integrals through a function $f: \mathbb{R}^2 \to \mathbb{R}$ is given by the Radon transform: 
\begin{eqnarray}
\mathcal{R} f  (\theta,s)= \int_{-\infty}^\infty f \left(s \boldsymbol \xi(\theta) + t \boldsymbol \xi^{\perp} (\theta) \right) \: dt, \label{E:radon} \\
\qquad \textrm{where } \boldsymbol \xi(\theta) = (\cos \theta, \sin \theta)^T, \: \boldsymbol \xi^\perp(\theta) = (-\sin \theta, \cos \theta)^T. \nonumber
\end{eqnarray}  

This transform is frequently used to model X-ray data. Assuming $f$ is compactly supported on the unit disc, it can be recovered exactly from $\mathcal{R}f$ if the latter is known for all $\theta \in [-\pi/2, \pi/2)$ and $s \in [-1, 1]$. In practice one only measures X-ray data for a discrete set of values $\theta_j$ and $s_l$; for a parallel-beam geometry with equally-spaced views and rays measured over a 180$^\circ$ arc, these are given by:
\begin{eqnarray}
 \theta_j = \pi(j-1)/p, \qquad & j = 1,2, \dots p, \label{E:sampling1}\\
s_l = l/q, & l = -q,-q+1, \dots q \nonumber.%\label{E:sampling2}.
\end{eqnarray}
This discrete sampling of $\mathcal{R}f$ can be used to reconstruct an image of $f$ accurately, provided that the rate of sampling is adequate. In particular, if $f$ is $b$-band limited (i.e., its Fourier transform vanishes for frequencies higher than $b$), then it is sufficient to take $p \geq b$ and $q \geq b/ \pi$~\cite{N86} (p. 71). Furthermore, to reconstruct an $n \times n$ pixel image of $f$, one can simply take $p = \lceil n \pi  / 2 \rceil$ and $q = \lceil n/2 \rceil$, since features with frequency higher than $n\pi/2$ cannot be represented at that resolution. This corresponds to acquiring $p$ views of the object equally spaced over $[-90^\circ,90^\circ)$ -- or equivalently, $[0^\circ, 180^\circ)$ -- with $n+1$ measurements per view\footnote{To distinguish between the two, we will use radians to refer to the parameter $\theta$ in \Eref{E:radon}, and degrees when referring to views of the object, which are acquired orthogonally to $\theta$.}. This choice of $p$ is conservative, as the true value of $b$ may be less than $n\pi / 2$. 

\begin{figure}
\includegraphics[width=\linewidth]{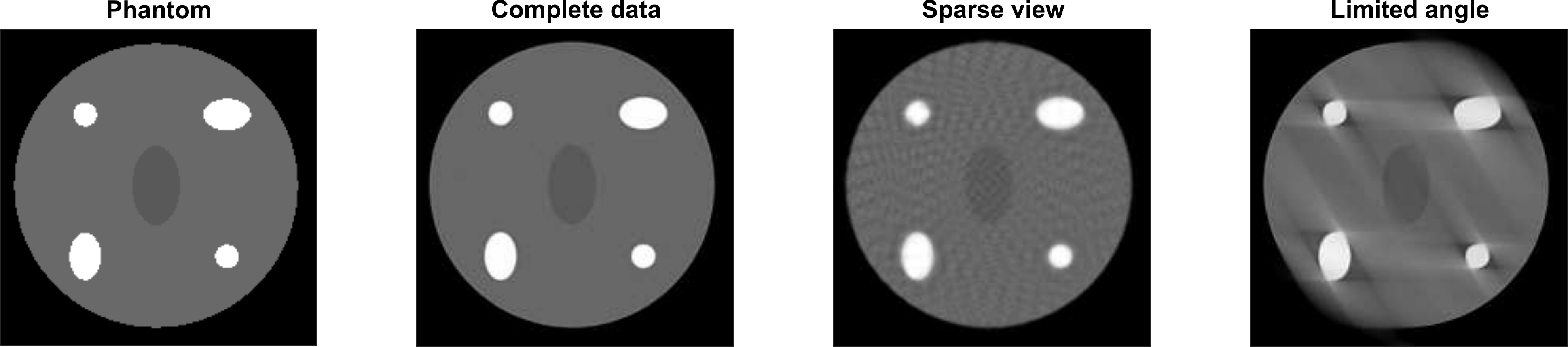}
\caption{Numerical phantom experiment showing artifacts characteristic of sparse-view and limited-angle reconstructions. All reconstructions are generated from monoenergetic data using SART. Left: Discretized phantom at resolution of $200 \times 200$ pixels. Centre left: reconstruction from complete data consisting of 360 equally spaced views on $[0^\circ,180^\circ)$. Centre right: sparse-view reconstruction from 36 equally spaced views on $[0^\circ,180^\circ)$. Right: limited-angle reconstruction from 240 equally spaced views on $[0^\circ,120^\circ)$.  }\label{F:artifacts}
% Phantom is 200 $\times$ 200 pixels with pixel size of 0.15 cm, and includes attenuation volumes modeling fat (dark grey, $\mu = 0.178$ cm$^{-1}$), soft tissue (light grey, $\mu$=0.203 cm$^{-1}$) and bone ($\mu$=0.495 cm$^{-1}$) at reference energy of 70 keV.
\end{figure}

We can then define sparse-view imaging as the case where $p \ll n \pi/2$, and limited-angle tomography as the case where there exist indices $1 \leq j_a < j_b \leq p$ such that either 
\begin{eqnarray}
j = j_a, j_a + 1, \dots j_b, \textrm{ or} \label{E:limitedangle} \label{E:sampling2} \\ 
j = 1, 2, \dots j_a, j_b, j_b +1, \dots  p, \nonumber 
\end{eqnarray}
in \eref{E:sampling1}, i.e., where some portion of the angular range is missing. \Fref{F:artifacts} shows some characteristic artifacts that arise in these two cases. For this 200$\times$200 pixel phantom, we have $n\pi/2 = 314$, which we round up to the more convenient value of $p=360$ to define complete data. The sparse-view reconstruction (centre right image) uses $p=36$, producing high-frequency artifacts characteristic of this case. The limited-angle reconstruction (right image) uses $p=360$ but $j = 1, \dots 60, 181, \dots 3600$, corresponding to 240 views acquired over $[0^\circ,120^\circ)$. In this case the artifacts consist of lost edges and streaks whose orientations are determined by the missing values of $\theta$. Specifically, one sees that edges are lost corresponding to the missing range of angles $\theta \in \left[ \pi / 6, \pi / 2\right)$, when there are no measurements acquired along lines tangent to those edges. These artifacts are characteristic of limited-angle reconstructions~\cite{Q93,FQ13}. 

\subsection{Mathematical head phantom}
\begin{figure}
\begin{center}
\begin{tabular}{cccc}
\includegraphics[width=0.2\linewidth]{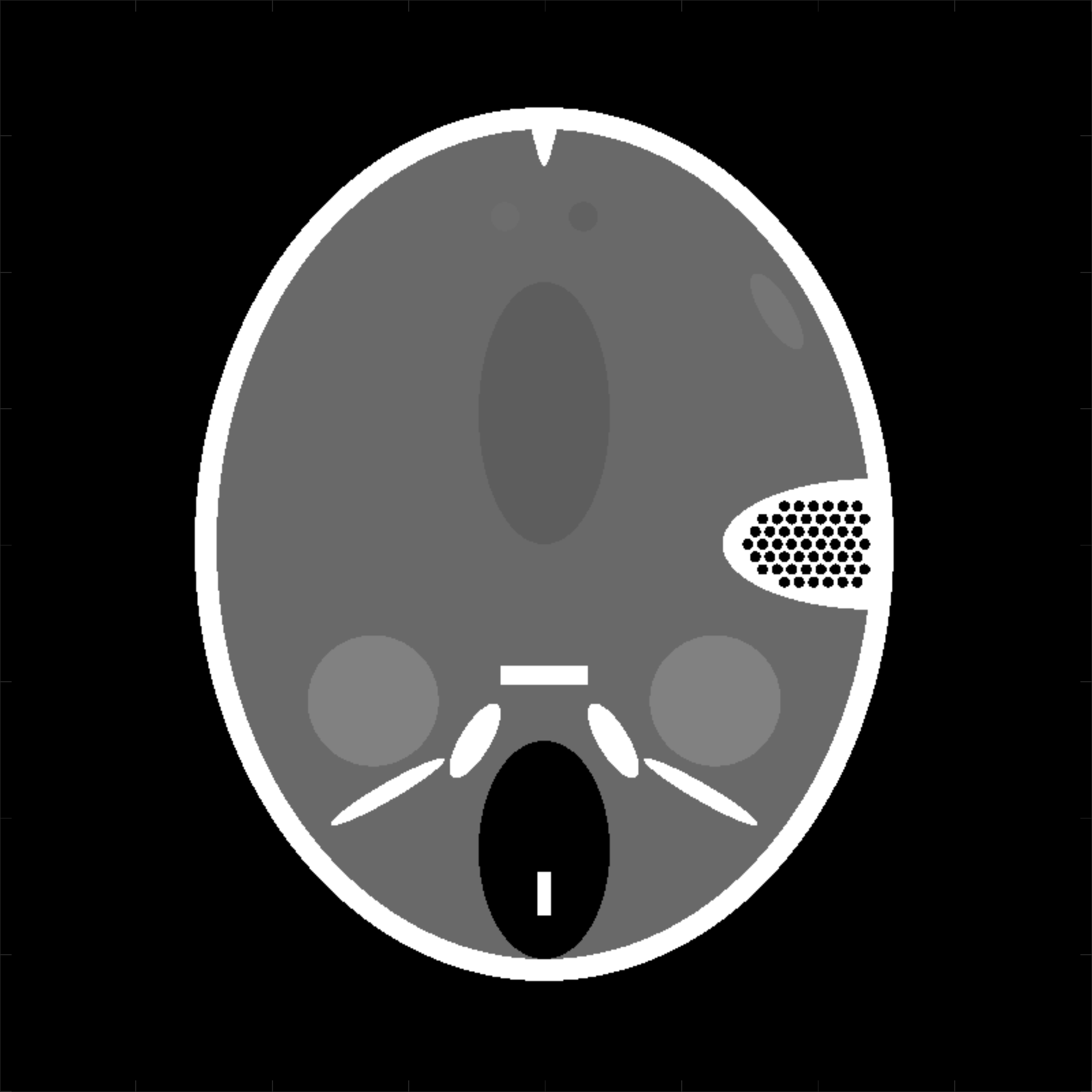} &\includegraphics[height=0.2\linewidth]{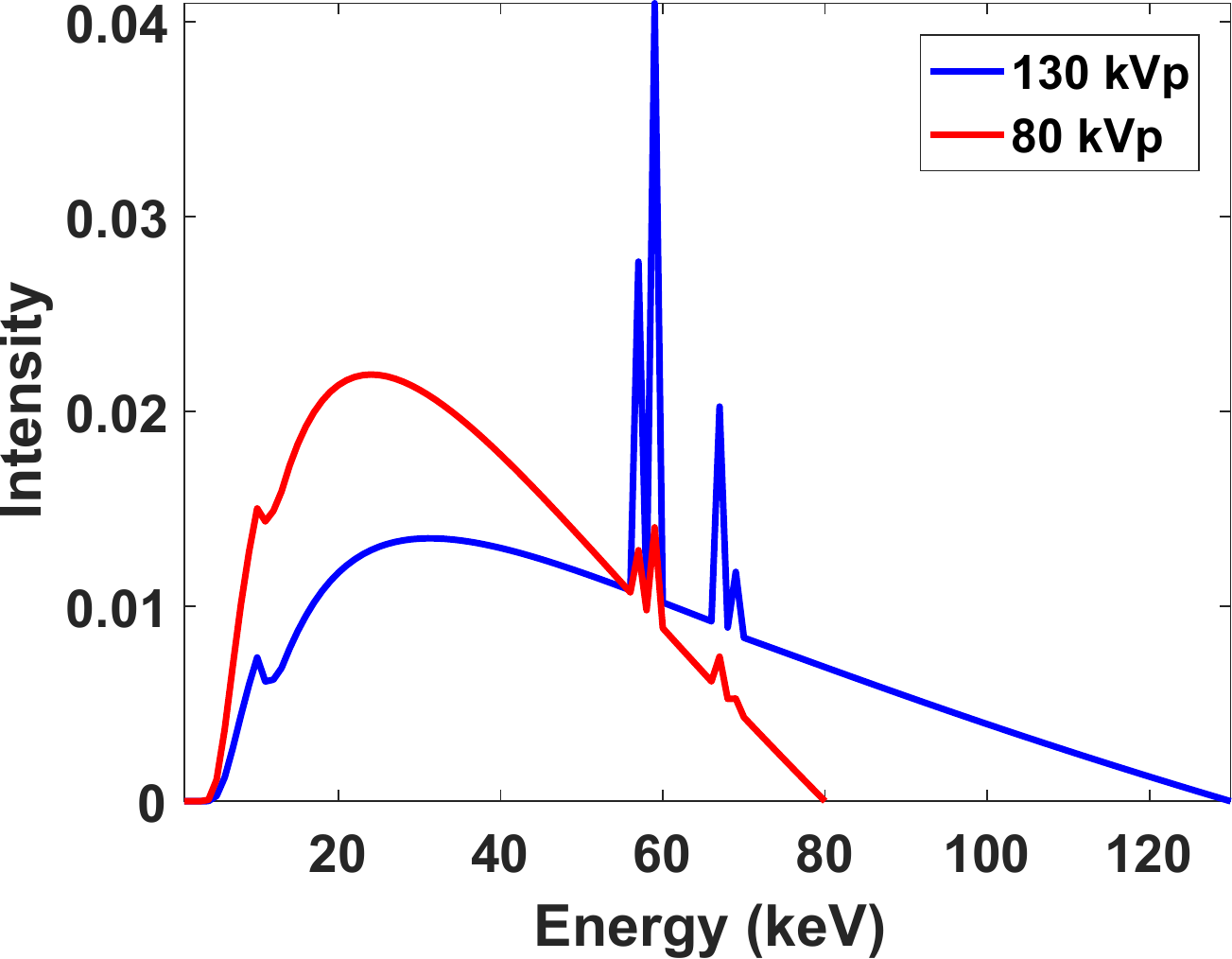} &\includegraphics[width=0.2\linewidth]{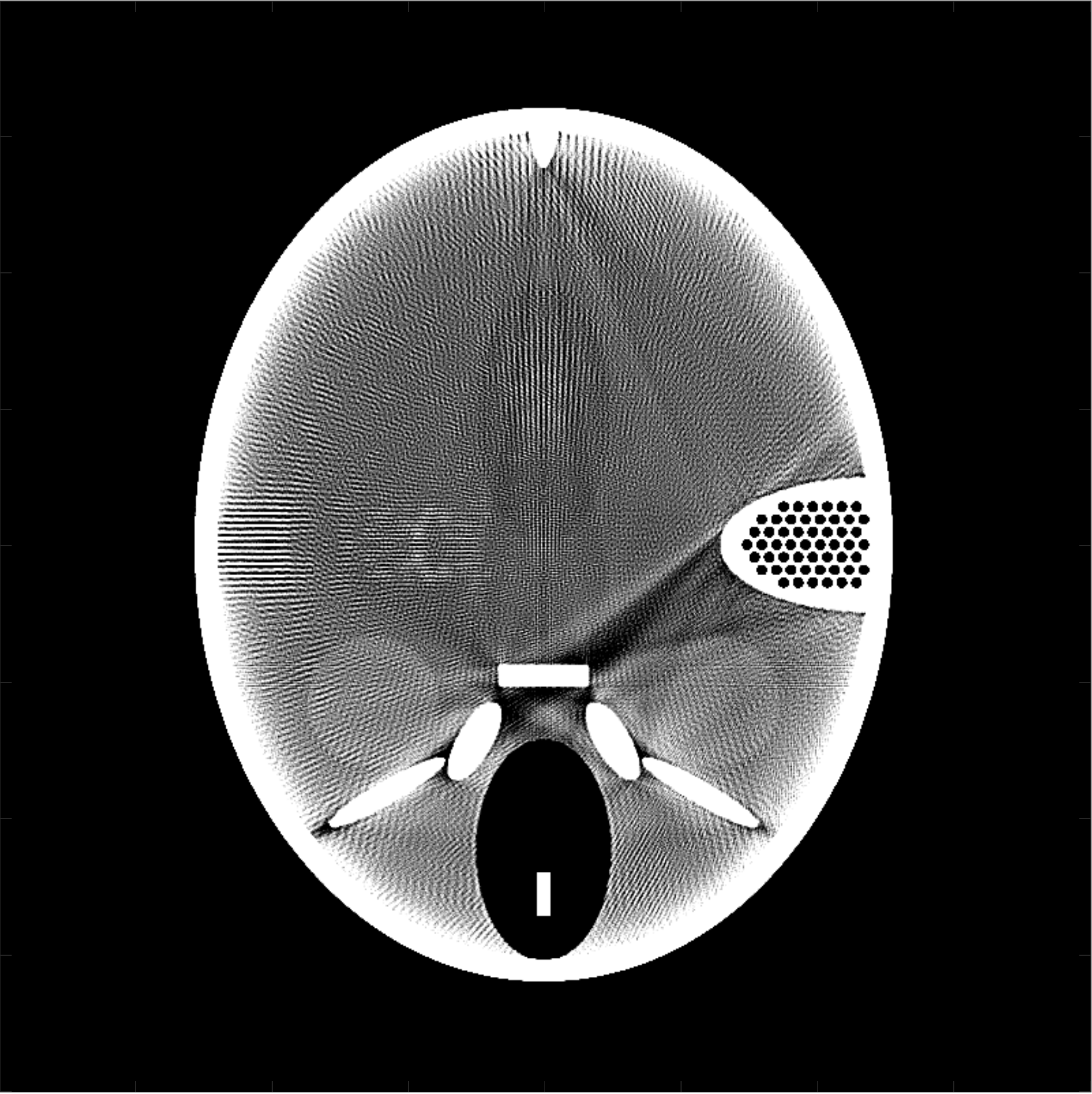} &\includegraphics[width=0.2\linewidth]{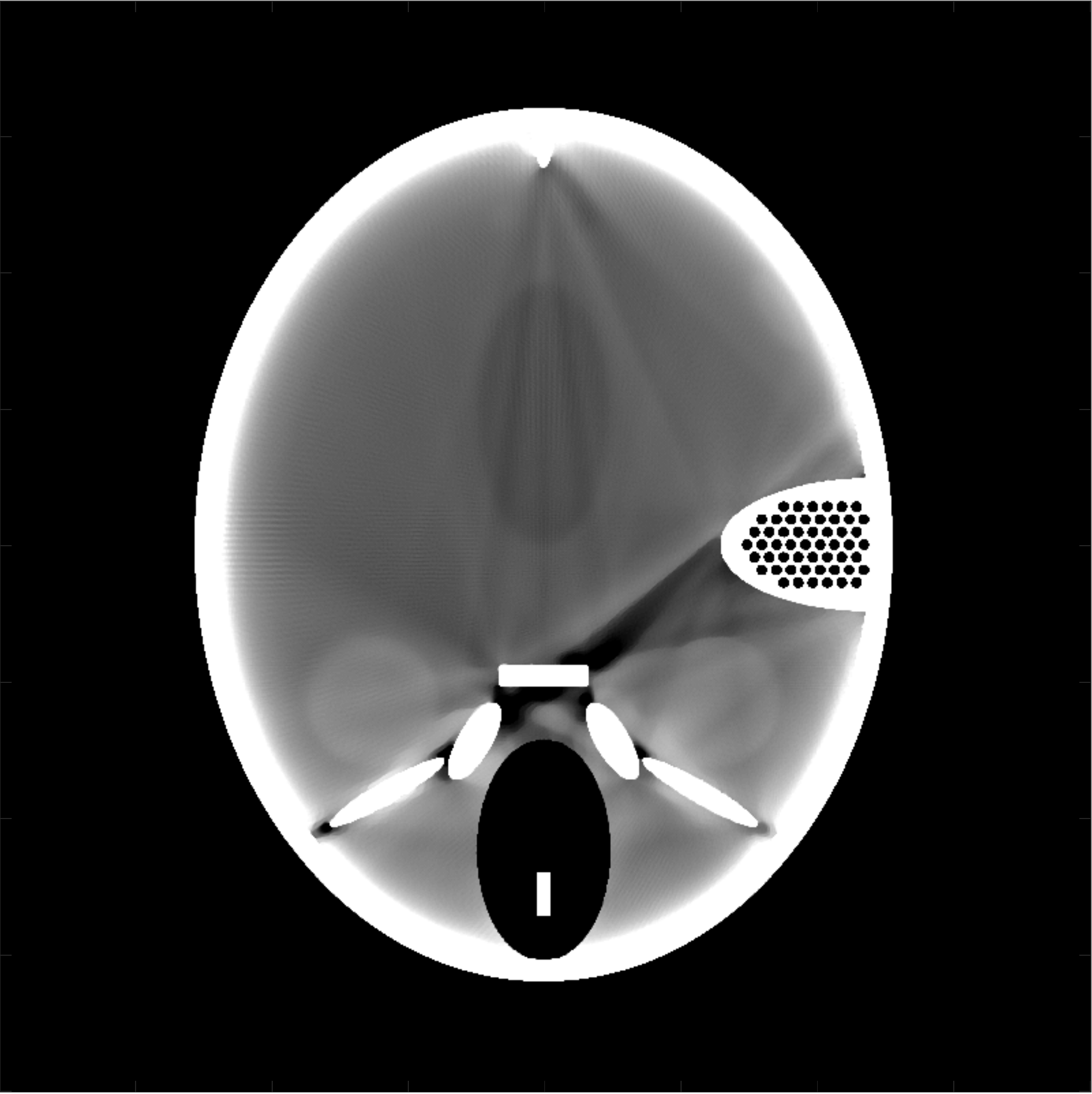} 
\end{tabular}
\end{center}
\caption{Left: FORBILD phantom at reference energy of 70 keV. Centre left: Normalized 130 kVp and 80 kVp spectra used to generate polyenergetic data.  Centre right: Image reconstructed using from soft tissue corrected data using SART. Right: image reconstructed from soft tissue corrected data using TV-superiorized SART.}\label{F:FORBILD}
\end{figure}

Our first set of numerical experiments use the FORBILD head phantom~\cite{YNDW12}. The phantom at a reference energy of 70 keV is shown in Fig.~\ref{F:FORBILD}. The intensity of the ROIs are all within $\pm 1\%$  of the background soft tissue intensity of 0.203 cm$^{-1}$, so all images are displayed on a narrow greyscale window of $[0.195, 0.215]$ cm$^{-1}$. The attenuation coefficient of the bone (white regions) at 70 keV is 0.495 cm$^{-1}$. The phantom size is $800 \times 800$ pixels with a pixel width of 0.375 mm. %This phantom simulates a 2D slice through the head, and includes an ear insert on the right side of the head as well as six low contrast regions of interest (ROIs). 

We generated several data sets from this phantom. The first set consisted of ``consistent'' projection data generated from the discrete phantom, using the same forward model assumed by pSART~\eref{E:project_poly},~\eref{E:mu_interp}. Thus the true phantom represented the exact solution to the nonlinear system of equations~\eref{E:project_poly}. This situation is unrealistic (an example of the ``inverse crime'') but is useful for examining the performance of the algorithm in the ideal case. The 130 kVp spectrum is shown in \Fref{F:FORBILD} and was generated using the Michigan Image Reconstruction Toolbox (MIRT)~\cite{Fessler}. We refer to this consistent data set as Set C.

We also generated three inconsistent data sets to test the performance of the approach under more realistic circumstances. For all three data sets, projections were generated using analytically computed line integrals through the FORBILD phantom, and Poisson-distributed noisy measurements were generated from the analytic data. Additionally, we introduced error into the spectrum used for reconstruction, by computing an approximation using the composite trapezoid rule. The parameters for these three data sets were:
\begin{itemize}
\itemsep0em
\item (Set I-1) 130 kVp spectrum with noise proportional to $\displaystyle \int I_0(E) \: dE = 4 \times 10^6$.
\item (Set I-2) 130 kVp spectrum with $\displaystyle \int I_0(E) \: dE = 1 \times 10^6$.
\item (Set I-3) 80 kVp spectrum (see \Fref{F:FORBILD}) with $\displaystyle \int I_0(E) \: dE = 4 \times 10^6$. 
\end{itemize}
Sets I-2 and I-3 are more challenging to accurately reconstruct than Set I-1, due to the higher level of noise in both datasets and lower average energy of the spectrum in Set I-3, which exacerbates beam hardening artifacts. The use of a low energy spectrum of this type is common in dual-energy CT imaging.%It is reasonable to assume, however, that an 80 kVp spectrum might be used in a dual-energy CT scan.

%We generated two sets of data from this phantom. The first set consisted of ``consistent'' projection data generated from the discrete phantom, using the same forward model assumed by pSART. In particular, \Eref{E:mu_interp} was used to generate the attenuation map of the phantom at all 130 energy levels in a 130 kVp spectrum, and projection data were generated by multiplying each of these attenuation maps by the system matrix $A$ and weighting it with the appropriate spectrum weight $S_h$ in~\Eref{E:project_poly}. Thus the true phantom represented the exact solution to the nonlinear system of equations~\eref{E:project_poly}. This situation is unrealistic (an example of the ``inverse crime'') but is useful for examining the performance of the algorithm in the ideal case. The spectrum is shown in \Fref{F:FORBILD} and was generated using the Matlab Image Reconstruction Toolbox~\cite{Fessler}.

%For the second data set, projections were generated using analytically computed line integrals through the FORBILD phantom, and Poisson-distributed noisy measurements were generated assuming an initial intensity of $\displaystyle \int I_0(E) \: dE = 4 \times 10^6$. Furthermore, the spectrum used for reconstruction was an approximation to the true spectrum, consisting of only 23 discrete energy levels with spectrum weights calculated using the composite trapezoid rule. This ``inconsistent'' data set therefore modeled the more realistic case where there is a discrepancy between the data and the forward model used for reconstruction.

As discussed in the previous section, reconstructing an $800 \times 800$ pixel image requires (conservatively) $400 \pi \approx 1250$ views acquired over 180$^\circ$. For convenience, we took $p=1440$ to obtain an integer multiple of 360, and defined this choice along with $j = 1, 2, \dots p$ to be the ``complete'' data set for both the consistent and inconsistent data. Our sparse-view and limited-angle experiments were then obtained by reducing $p$ and/or taking a subset of $j$ values as described in \Eref{E:sampling2}. 

Before presenting the results of our experiments, we provide some motivation for superiorizing a polyenergetic reconstruction algorithm such as pSART. One possible approach to reconstructing an image from sparse-view or limited-angle polyenergetic data would be to generate soft tissue corrected data and then use a superiorized version of a monoenergetic algorithm, such as SART. Soft tissue correction~\cite{JS78,HT83} is a simple procedure in which monoenergetic data is simulated from polyenergetic data using the following steps:
\begin{enumerate}
\item Solve  $\displaystyle I_\ell = \sum_{h=1}^{N_h} S_h \exp \left(-\mu_s(E_h)T_{\ell} \right)$ for $T_\ell$, for all measurements $\ell$.
\item Let $m_\ell = I_0 \exp \left(-\mu_s(E_0\right) T_\ell)$.
\end{enumerate}
Here $\mu_s(E)$ denotes the attenuation coefficient of soft tissue at energy $E$. The first step consists of solving a nonlinear equation for each measurement to determine the equivalent length, $T_\ell$, of soft tissue through which ray $\ell$ would have to pass to generate measurement $I_\ell$. The second step then generates the corresponding monoenergetic data at energy $E_0$, which can be reconstructed with a conventional algorithm, such as SART. While this technique is effective in lessening cupping artifacts, it does not remove streaking artifacts caused by high-attenuation material such as bone or contrast agents. \Fref{F:FORBILD} shows that this is true also for a TV-superiorized version of SART. The image in the centre right shows the FORBILD phantom reconstructed from 288 soft tissue corrected views using SART, while the image on the right shows a reconstruction from the same data using TV-superiorized SART. While the TV minimization is effective in removing high-frequency artifacts caused by the sparse-view data, it does not remove the streaking artifacts caused by beam hardening. While techniques similar to soft tissue correction can be extended to account for artifacts of this nature~\cite{JS78,JR97}, they typically require image segmentation and do not account for mixtures of different tissue types within pixels. By superiorizing pSART we avoid both of these issues.

%\begin{figure}
%\centering
%\begin{tabular}{cc}
%
%\end{tabular}
%\caption{Images reconstructed from soft-tissue corrected polyenergetic sparse-view data.}\label{F:recon_wc}
%\end{figure}

\subsection{Sparse-view experiments}

In addition to the complete data set with $p=1440$, we reconstructed data using $p =  720$, 360, 288, 144 and 72. Images were reconstructed using both pSART and pSART-TV, beginning with an initial guess of $\xvec^{(0)} = \mathbf{0}$ in all cases, using a reference energy of 70 keV for data sets C, I-1 and I-2, and 50 keV for data set I-3, due to the lower average energy of the spectrum. Air, soft tissue, and bone were used as the basis materials, with the energy-dependent attenuation curves obtained from the MIRT. The number of subsets, $N_w$, was chosen such that every subset consisted of 12 views; this corresponds to $N_w = 120$ for $p=1440$, down to $N_w = 6$ for $p=72$. pSART-TV was run with $\gamma = 0.999$ and $N = 20$ as the counter for the TV minimization loop.  %and titanium\footnote{The phantom itself does not contain titanium, but including an attenuation value greater than bone is useful to perform interpolation, in case the values in some pixels happen to exceed the bone value at some point during the reconstruction.} , corresponding to one half, one quarter, one fifth, one tenth and one twentieth the number of views, respectively

As discussed in Section~\ref{S:pSARTsup}, the parameter $\varepsilon_{target}$ should correspond to a solution produced by the unsuperiorized algorithm, pSART. This leads to the question of when pSART should be terminated to generate this solution. In our experiments we chose to terminate the algorithm after a fixed number of iterations were run. This was the approach used in~\cite{LS14b} and has the advantage of being simple, although the total number of iterations to run must be determined ad-hoc. When using inconsistent data we ran fewer iterations of pSART since it exhibits semiconvergence behaviour similar to SART and other methods~(see e.g.~\cite{EHN14}). Semiconvergence refers to the degradation of image quality after a certain number of iterations, as the algorithm begins fitting noise and other inconsistencies in the data. In both the consistent and inconsistent data case the total number of iterations to run was increased as $p$ decreased, since the cost per iteration decreases proportionately to the number of views. This ensured that the same amount of computation (in terms of forward projection operations) was done for all values of~$p$. 

%For inconsistent data, it is well known that the accuracy of an image produced by algorithms like ART eventually begins to worsen (even as the residual is improved) since after a certain point the algorithm simply fits inconsistencies in the data, such as noise~\cite{EHN14}. In this case, we chose to terminate pSART after the first iteration in which the image error (measured as $\norm{\xvec^{(k)}-\xvec^{true}}_2$) was larger than the previous iteration. This stopping criteria can obviously not be used in practice when the true image, $\xvec^{true}$, is unknown. We found it to be a useful criterion in this experiment, however, as it ensured that pSART-TV was finding a solution that was as constraints-compatible as the best solution found by pSART. In general we found that this termination criterion resulted in running fewer iterations than we chose to run for the consistent data set.

\begin{figure}
\includegraphics[width=\linewidth]{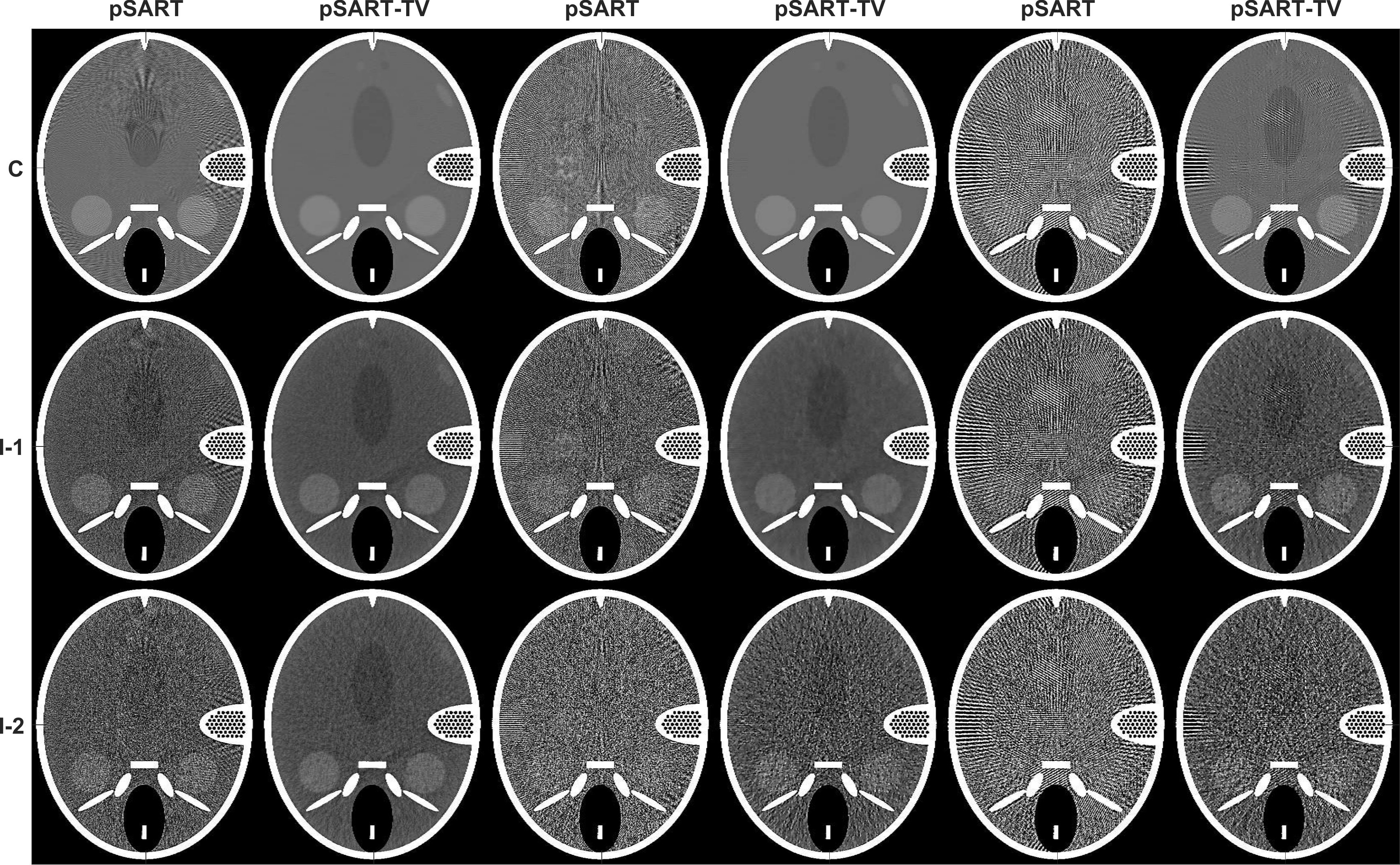}\\[-2px]
\includegraphics[width=\linewidth]{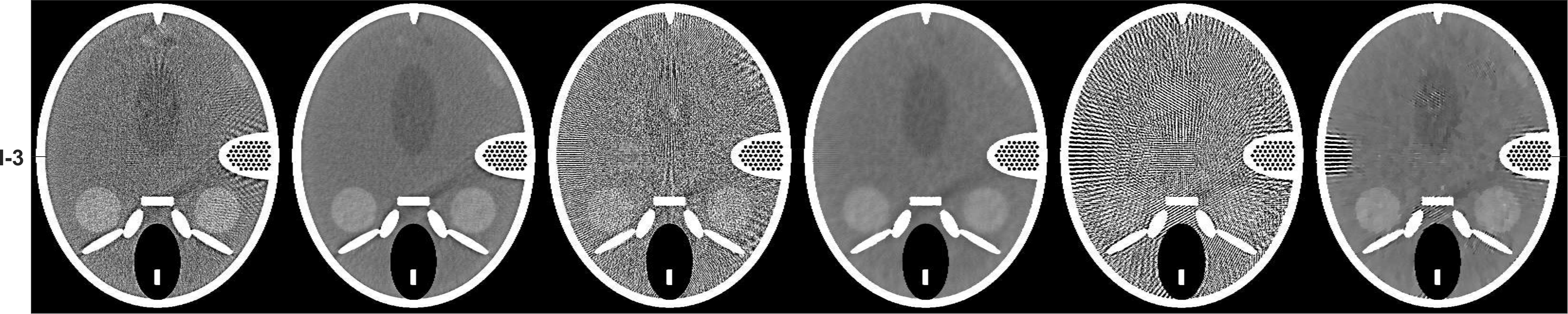}

\begin{tabular}{p{0.33\linewidth}p{0.33\linewidth}p{0.33\linewidth}} \centering $p = 1440$ & \centering{ $p = 288$} & \centering{$p = 144$}
\end{tabular}
%\begin{tabular}{ccc}
%\includegraphics[width=0.25\linewidth]{fig5a.pdf} &\includegraphics[width=0.25\linewidth]{fig5b.pdf} &\includegraphics[width=0.25\linewidth]{fig5c.pdf} \\
%\includegraphics[width=0.25\linewidth]{fig5d.pdf} &\includegraphics[width=0.25\linewidth]{fig5e.pdf} &\includegraphics[width=0.25\linewidth]{fig5f.pdf} \\
%$p = 1440$ &$p = 288$ &$p = 144$
%\end{tabular}
\caption{Images of FORBILD phantom reconstructed from sparse-view data using pSART and pSART-TV. From top to bottom, rows correspond to images reconstructed from data sets C, I-1, I-2 and I-3, respectively. First, third and fifth columns show images reconstructed with pSART, while second, fourth and sixth columns show corresponding images reconstructed with pSART-TV. Number of views used for each column pair is indicated at the bottom.}\label{F:sparseview_images}
\end{figure}

Some representative images are shown in \Fref{F:sparseview_images}.  In the consistent data case (top row), the image reconstructed by pSART-TV using $p=288$ (fourth column) is virtually indistinguishable from the image obtained with $p=1440$ (second column), though there is a slight artifact on the left side of the phantom opposite the ear insert. This artifact becomes more pronounced when $p$ is reduced further to 144 views (sixth column), with additional artifacts also arising. Due to the polyenergetic forward model, the images are virtually free of any beam hardening artifacts (cf. \Fref{F:FORBILD}), although the streak caused by the ear insert is faintly visible.

For data sets I-1, I-2 and I-3, the image quality is diminished due to the various sources of inconsistency. The reconstructed intensity values from data set I-1 (second row) are too low due to the mismatch in the spectrum, and the large streak caused by the ear insert is somewhat more prominent than for the consistent data. The addition of noise also causes the images to have a patchier appearance, and makes it more difficult to discern the three small low-contrast ROIs near the edge of the skull, especially as $p$ decreases. When the noise level increases in Set I-2, this problem is exacerbated and some of the low-contrast features are not apparent in the image. Finally, in the images reconstructed from Set I-3 (which are displayed using a greyscale window of $[0.235, 0.255]$ cm$^{-1}$ due to the lower reference energy), the noise level is not as high as in Set 1-B, but the beam hardening artifacts are somewhat more noticeable, as one might expect from a lower energy spectrum. %In particular, the mismatch in the spectrum causes the reconstructed intensity values to be too low, and the large streak caused by the ear insert is somewhat more prominent than for the consistent data. The addition of noise also causes the images to have a patchier appearance, and makes it more difficult to discern the three small low-contrast ROIs near the edge of the skull, especially as $p$ decreases. As in the consistent data case, artifacts due to the undersampling become much more prominent when reducing $p$ from 288 to 144.%We note that the undersampling artifacts for the case $p=144$ (bottom right) appear less severe than when consistent data was used (top right); this is also reflected in their respective TV values of 2551 and 3268. This discrepancy is due to the fact that pSART-TV was run for 227 iterations in the consistent data case and only 108 iterations in the inconsistent case, due to the different stopping criteria that were used. The undersampling artifacts become more prominent as further iterations are run with inconsistent data as well.

\begin{figure}
\begin{tabular}{ccc}
\includegraphics[height=0.2\linewidth]{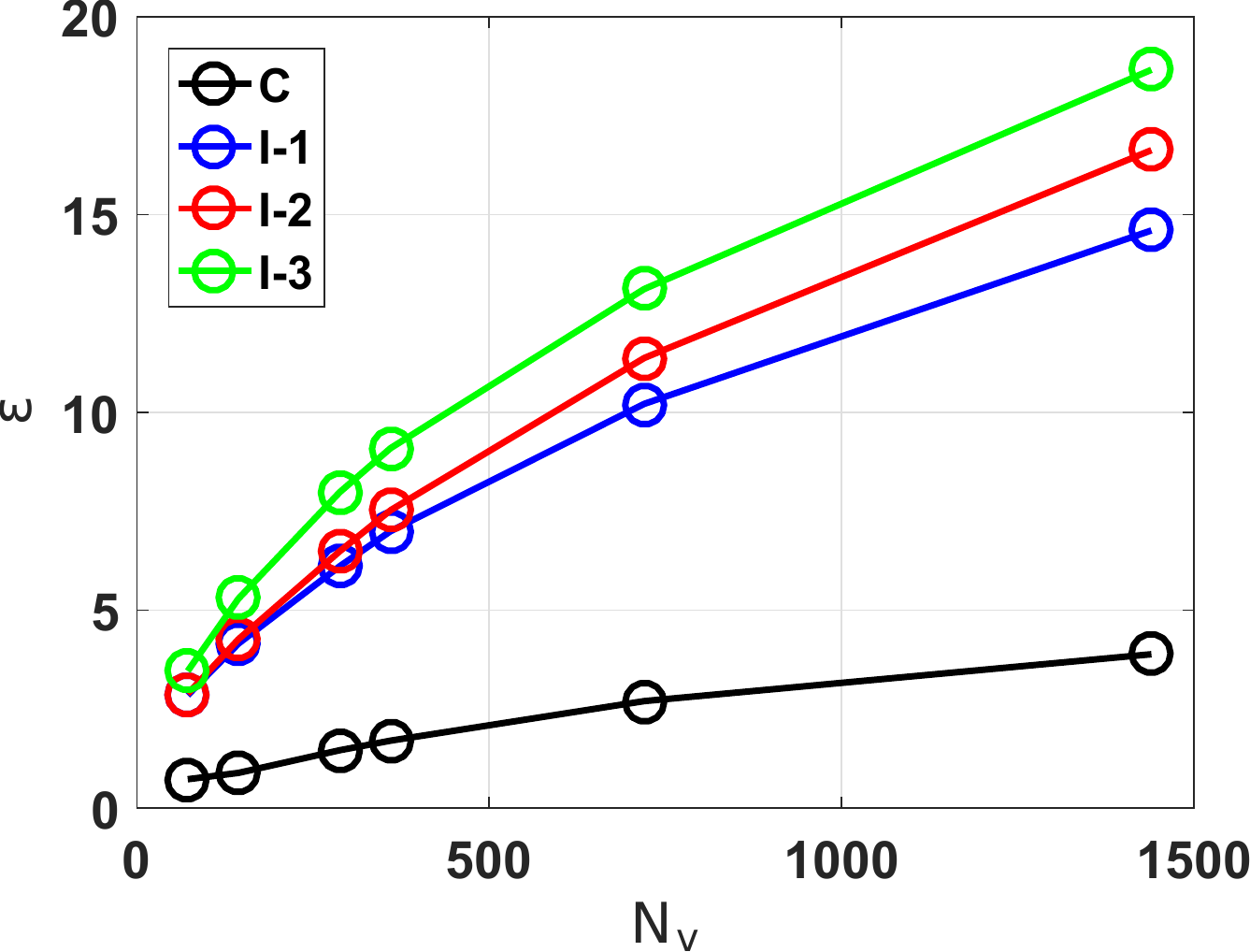} &\includegraphics[height=0.2\linewidth]{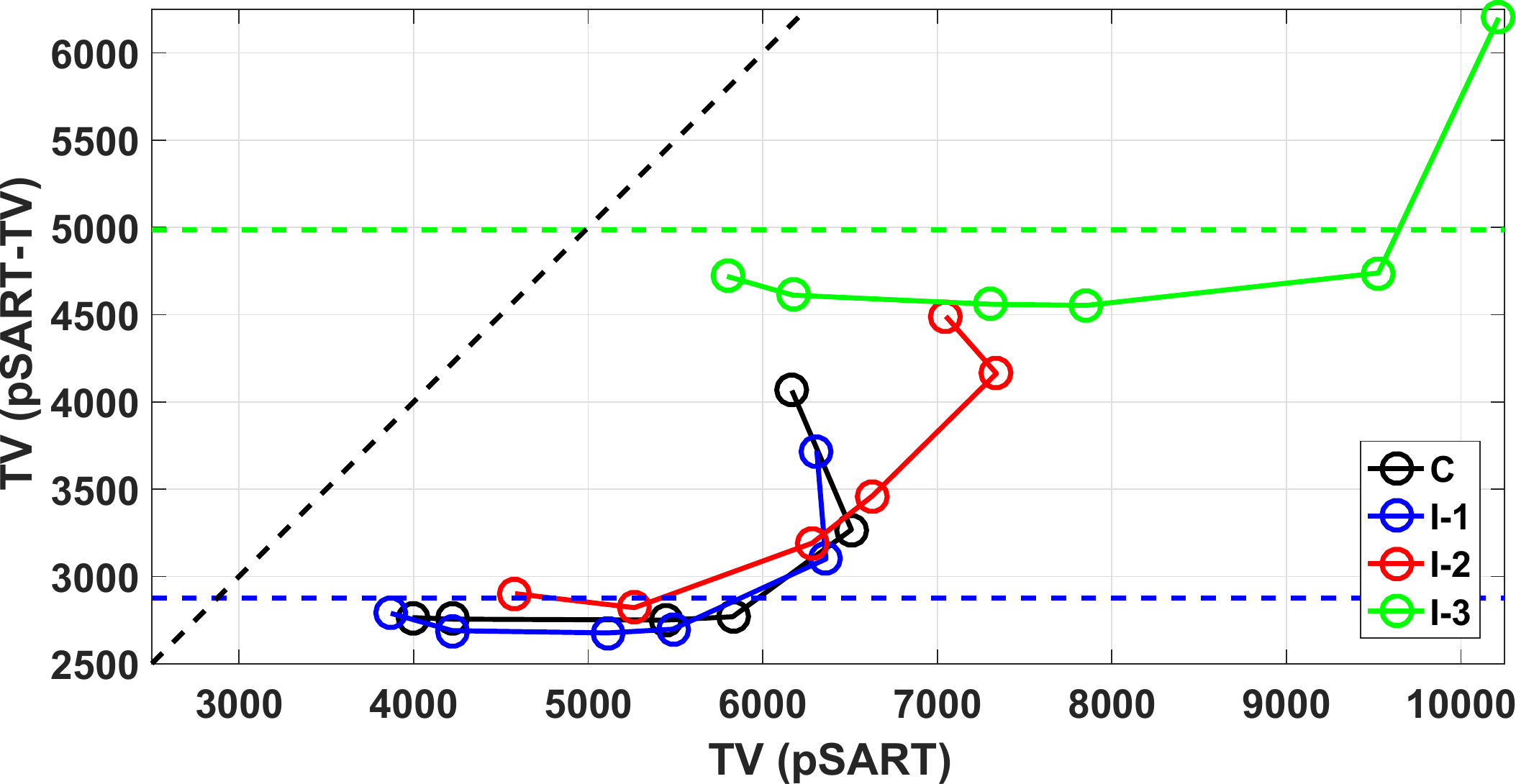} &\includegraphics[height=0.2\linewidth]{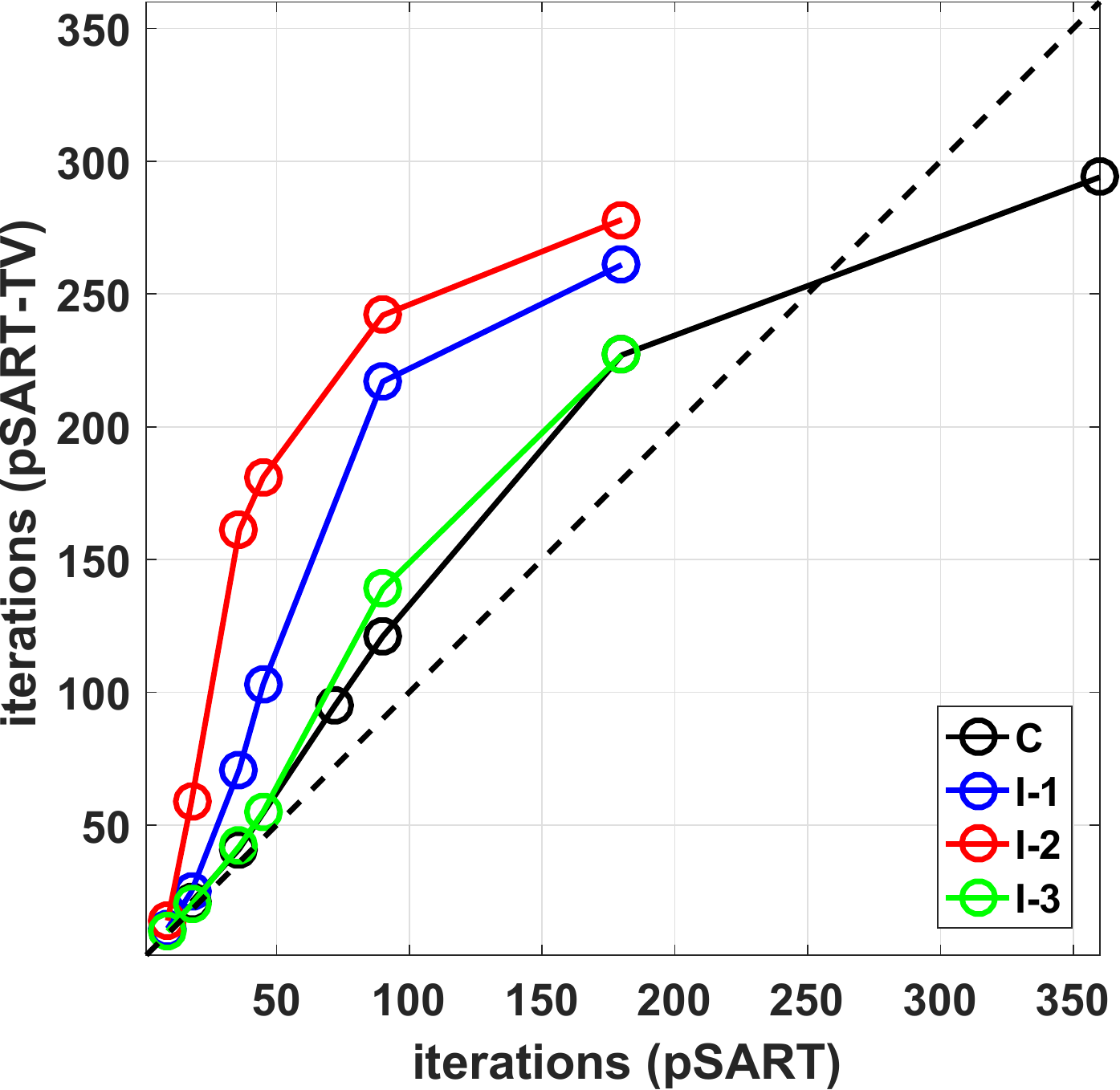}
\end{tabular}
\caption{Plots of important quantities from sparse-view experiments. Left plot: Values of $\varepsilon$ obtained by pSART and pSART-TV for each simulation. Middle plot: TV values for pSART reconstructions plotted against TV values for pSART-TV reconstructions.Black dashed line is the line $y=x$, blue and green dashed lines represent the true TV of the phantom at 70 keV and 50 keV, respectively.  Leftmost points correspond to $p=1440$ in all cases, with $p$ decreasing as one follows the line. Right plot: number of iterations required by each algorithm for each simulation. Leftmost points correspond to $p=1440$, as before.}\label{F:sparseview_plots}
\end{figure}

While pSART-TV did not produce high quality images in all cases, the key claim of superiorization is simply that the superiorized algorithm should be able to find solutions that are as constraints-compatible as those found pSART, with lower TV. This was the case in every experiment that was run. Plots of relevant quantities are shown in \Fref{F:sparseview_plots}. The leftmost plot shows the $\varepsilon$ values obtained by both pSART and pSART-TV in every simulation, as a function of the number of views. The much smaller value of $\varepsilon$ for Set C (black curve) can be attributed to the fact that the data were consistent, and also that pSART was run for more iterations than in the inconsistent data cases. Additionally, since $\varepsilon$ is not normalized (see \eref{E:res_poly}), the values also decrease as fewer views are included. Importantly, the middle plot shows that the images found by pSART-TV had substantially smaller TV than the solutions found by pSART, as desired. One can also see that the TV values of the images reconstructed using pSART-TV were fairly consistent as $p$ decreased, and relatively close to the true TV values of the phantom at 70 keV (for the first three data sets) or 50 keV (for Set I-3). Finally, the rightmost plot shows the total number of iterations required by pSART and pSART-TV to achieve the same $\varepsilon$ value in each experiment. Unsurprisingly, pSART-TV required more iterations than pSART in nearly every case, as the perturbation introduced at every iteration competes with the goal of fitting the data (i.e. reducing $\varepsilon$). This was especially true for the noisy data sets.

%\begin{table}
%\caption{Results for sparse-view experiments. $N_v$ refers to number of views used for reconstruction (equal to $p$); next three columns correspond to images reconstructed from consistent data; last three columns to images reconstructed from inconsistent data. Values separated by a slash correspond to the pSART and pSART-TV reconstructions, respectively. $K$ refers to total number of iterations that were run, $\varepsilon$ to the value of the proximity function \eref{E:res_poly} obtained at termination for both algorithms, and TV to the total variation of the image \eref{E:TV1} obtained at termination.}\label{T:sparseview}
%
%\footnotesize
%\begin{tabular}{rrrrrrr}
%\\
%&\multicolumn{3}{c}{Consistent} &\multicolumn{3}{c}{Inconsistent} \\ 
%\multicolumn{1}{c}{$N_v$}  &\multicolumn{1}{c}{$K$} &\multicolumn{1}{c}{$\varepsilon$} &\multicolumn{1}{c}{TV} &\multicolumn{1}{c}{$K$} &\multicolumn{1}{c}{$\varepsilon$} &\multicolumn{1}{c}{TV} \\
%\cmidrule(r){1-1}\cmidrule(r){2-4} \cmidrule(r){5-7} 
%1440	 &18 /~~21 &3.890 &3993 / 2764 &9 /~~11 &14.60  &3869 / 2789  \\
%720	&36 /~~41 &2.700  &4218 / 2755&18 /~~25 &10.20  &4221 / 2689\\
%360	&72 /~~95 &1.700  &5442 / 2750&36 /~~71 &6.99  &5114 / 2676\\
%288	&90 /~121 &1.460  &5828 / 2769&45 /~103 &6.11  &5484 / 2697\\
%144	&180 /~227 &0.890  &6508 / 3268 &90 /~217 &4.16  &6361 / 3100\\
%72 	&360 /~294 &0.730  &6167 / 4065 &180 /~261 &2.85  &6310 / 3720
%\end{tabular}
%\end{table}

\subsection{Limited-angle experiments}

% Note: Below paragraph assumes that alpha and n_alpha were defined in section about the calculation of anistropic TV calculation.
To conduct limited-angle experiments, we reconstructed images from data acquired over angular extents of 165$^\circ$ and 150$^\circ$, corresponding to $1320$ and $1200$ total views, respectively. In all cases we used $p=1440$; for limited angle data this parameter is no longer equal to the total number of views, but rather defines the spacing between consecutive views as $180^\circ / p$, or $0.2^\circ$ in this case. For each angular extent, two different acquisition arcs were considered; one consisting of views acquired between $0^\circ$ and $180^\circ$, and the another consisting of views acquired between $90^\circ$ and $270^\circ$. For example, the two $165^\circ$ acquisitions consisted of views acquired on $[97.5^\circ, 262.5^\circ)$ and $[7.5^\circ, 172.5^\circ)$. With respect to Equations~\eref{E:sampling1} and \eref{E:sampling2}, these two cases correspond to  $j = 61, 62, \dots 1380$ and $j = 1, 2, \dots 660, 781, 782, \dots 1440$, respectively. For ease of reference, however, we will denote each acquisition arc by the starting angle (first view acquired) and the angular extent.

All images were reconstructed using both pSART and pSART-ATV with an initial guess of $\xvec^{(0)} = \mathbf{0}$. For the ATV calculation~\eref{E:ATV1}, we took $N_\alpha=4$ and $\alpha=\{0^\circ, 45^\circ, 90^\circ, 135^\circ\}$. As in the sparse-view case, the number of subsets $N_w$ was chosen such that every subset consisted of 12 views. Because the artifacts introduced by the limited-angle problem were, in general, more severe than those encountered in the sparse-view problem, we altered the pSART-ATV parameters to $\gamma = 0.9999$ and $N = 60$, which we found was more effective in lessening artifacts. As in the sparse-view experiments, we terminated pSART after a fixed number of iterations, increasing that number as the angular extent decreased, and running fewer iterations for the inconsistent data cases.

\begin{figure}
\includegraphics[width=\linewidth]{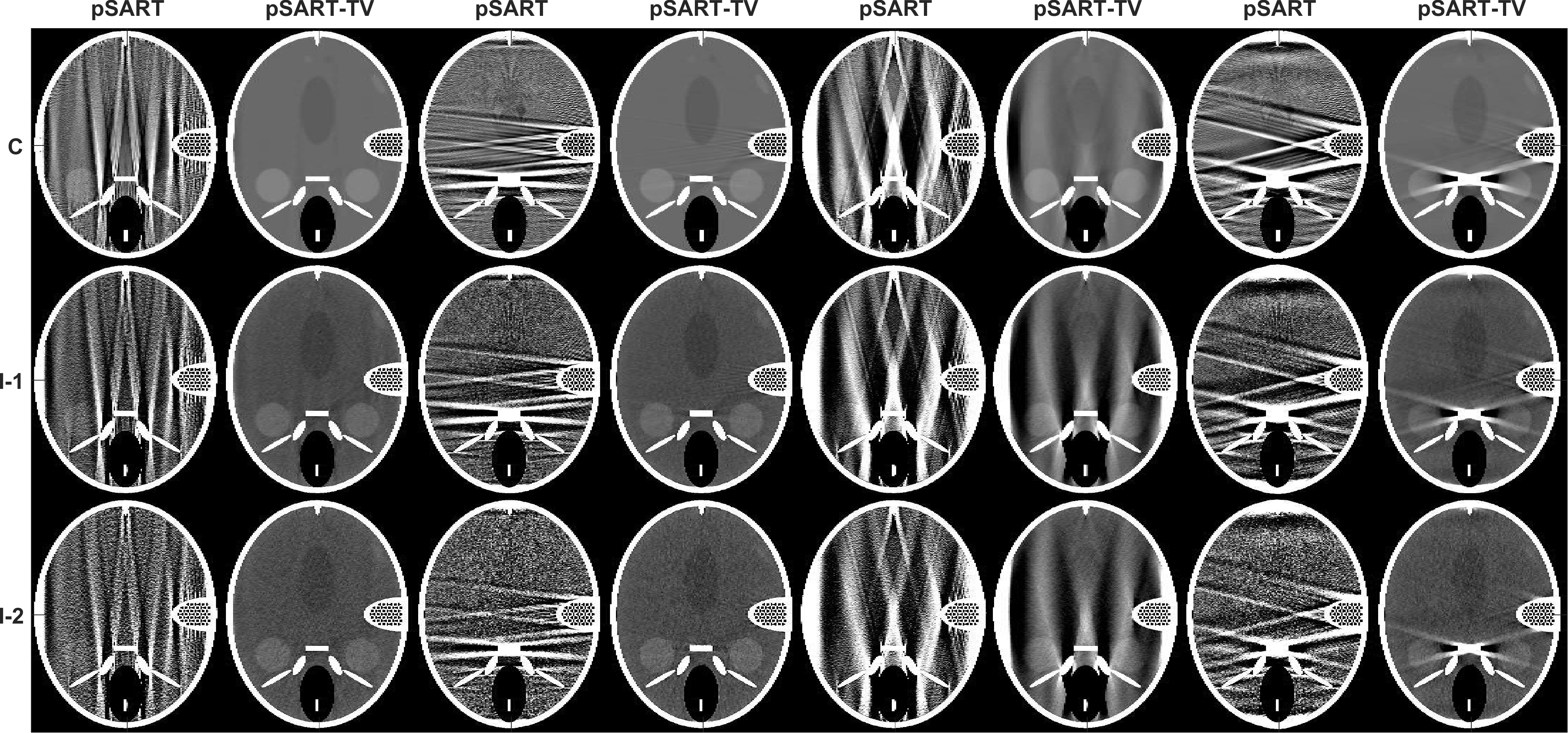}\\[-2px]
\includegraphics[width=\linewidth]{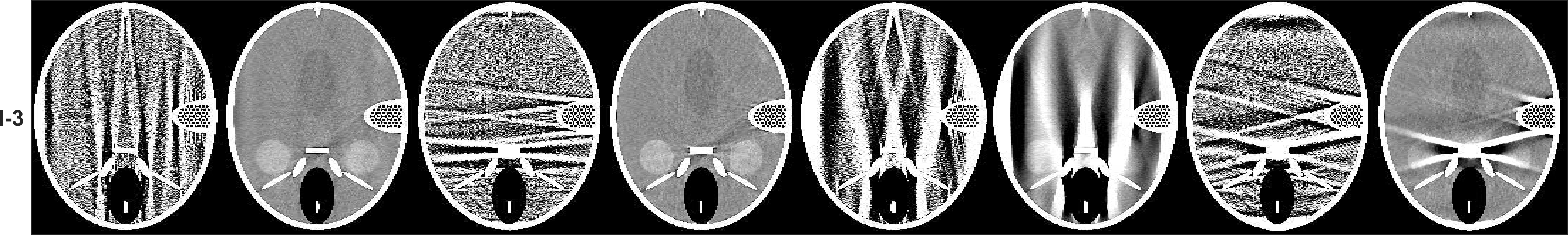}
	\begin{tabular}{p{0.25\linewidth}p{0.25\linewidth}p{0.25\linewidth}p{0.25\linewidth}}
		Extent: 165$^\circ$ & Extent: 165$^\circ$ & Extent: 150$^\circ$ & Extent: 150$^\circ$\\
		Start: $97.5^\circ$	& Start: $7.5^\circ$ & Start: $105.0^\circ$ & Start: $15.0^\circ$ 	
	\end{tabular}
	\caption{Images of FORBILD phantom reconstructed from limited angle data by pSART and pSART-ATV. Layout of rows and columns is the same as \Fref{F:sparseview_images}.}\label{F:limitedangleimages}
\end{figure}

Representative images of the limited-angle reconstructions are shown in \Fref{F:limitedangleimages}. The images obtained from the $165^\circ$ arcs and consistent data (top row, second and fourth columns) are generally accurate reconstructions of the phantom; some artifacts characteristic of limited-angle reconstructions are evident, but they are greatly reduced compared to the corresponding images reconstructed with pSART. When the starting view is $7.5^\circ$, the missing views correspond to the horizontal axis (and nearby angles) and so the streaks and lost edges occur along directions tangent to those missing views. When the starting view is $97.5^\circ$, the missing directions are close to the vertical axis and so the streaks occur along those directions, and are generally less severe since the phantom has fewer sharp edges in the vertical direction. When the angular extent is further reduced to $150^\circ$ (sixth and eighth columns), the artifacts become more severe and pSART-ATV is not able to remove them entirely. Artifacts caused by beam hardening, however, are only faintly visible (if at all).

Reconstructions from ithe inconsistent data sets are shown in the second, third and fourth rows. As in the sparse-view case, there is a loss of image quality due to the sources of inconsistency in the data; nonetheless, the superiorized algorithm remains effective at removing limited-angle artifacts. These artifacts appear to be no more severe than they were in the consistent data case, with the exception of the images in the sixth column (starting view of $105^\circ$). This can be attributed primarily to the fact that the image reconstructed from consistent data (top row) was produced from 128 iterations of pSART-ATV, while the images reconstructed from inconsistent data (second to fourth row) were produced from between 43--54 iterations. We discuss the relation between the stopping criterion for pSART and the number of iterations run by pSART-ATV later in this section. Aside from this, we note that as before, data sets I-2 and I-3 were more challenging to reconstruct than data set I-1; the images are visibly noisier, and the beam hardening artifact is more severe with data set I-3.

\begin{figure}
\begin{tabular}{ccc}
\includegraphics[height=0.2\linewidth]{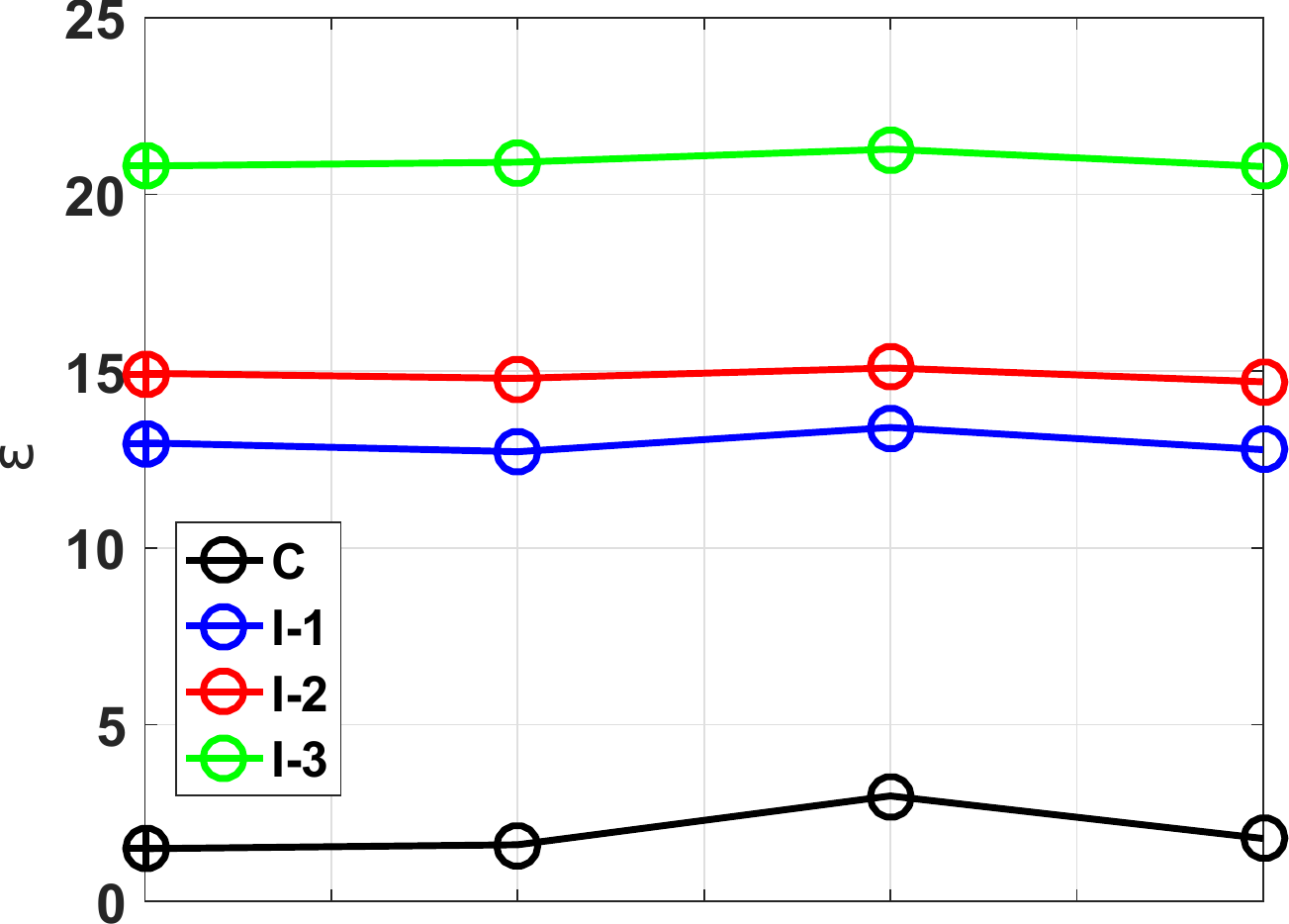} &\includegraphics[height=0.2\linewidth]{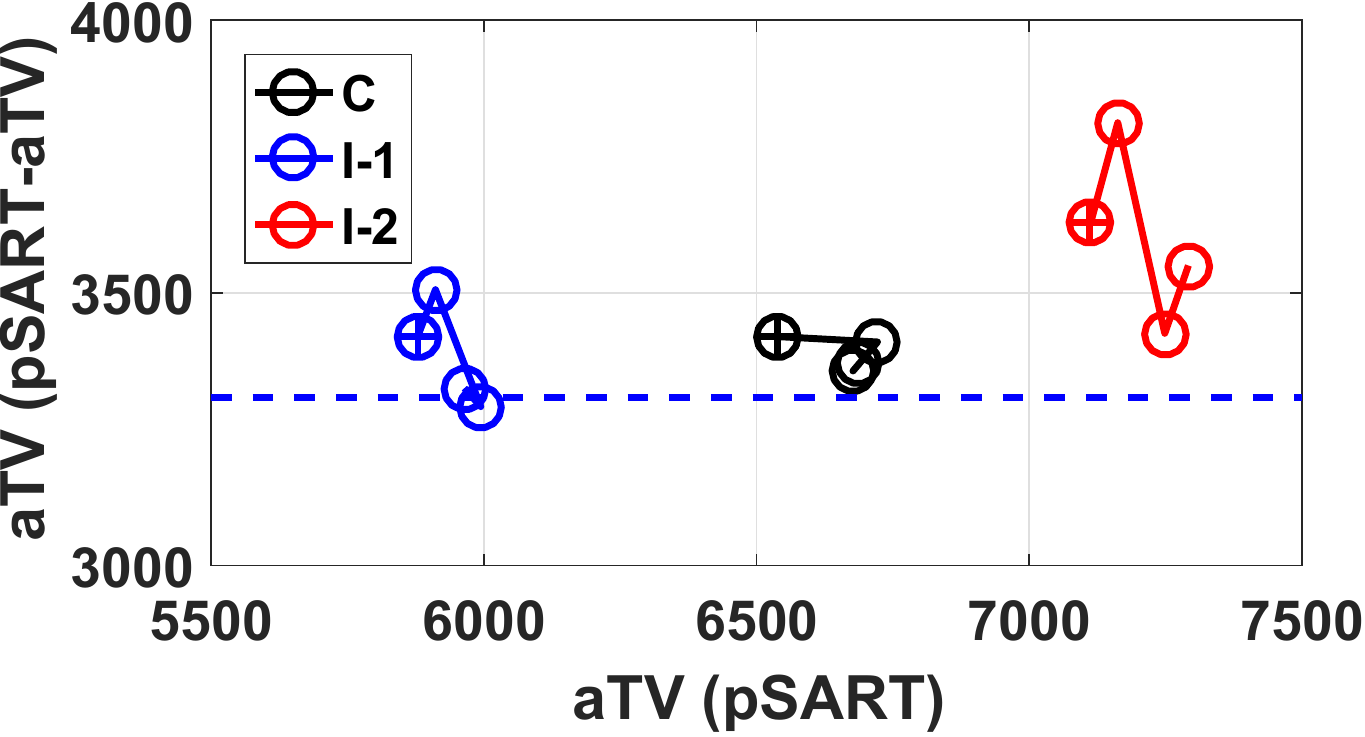} &\includegraphics[width=0.2\linewidth]{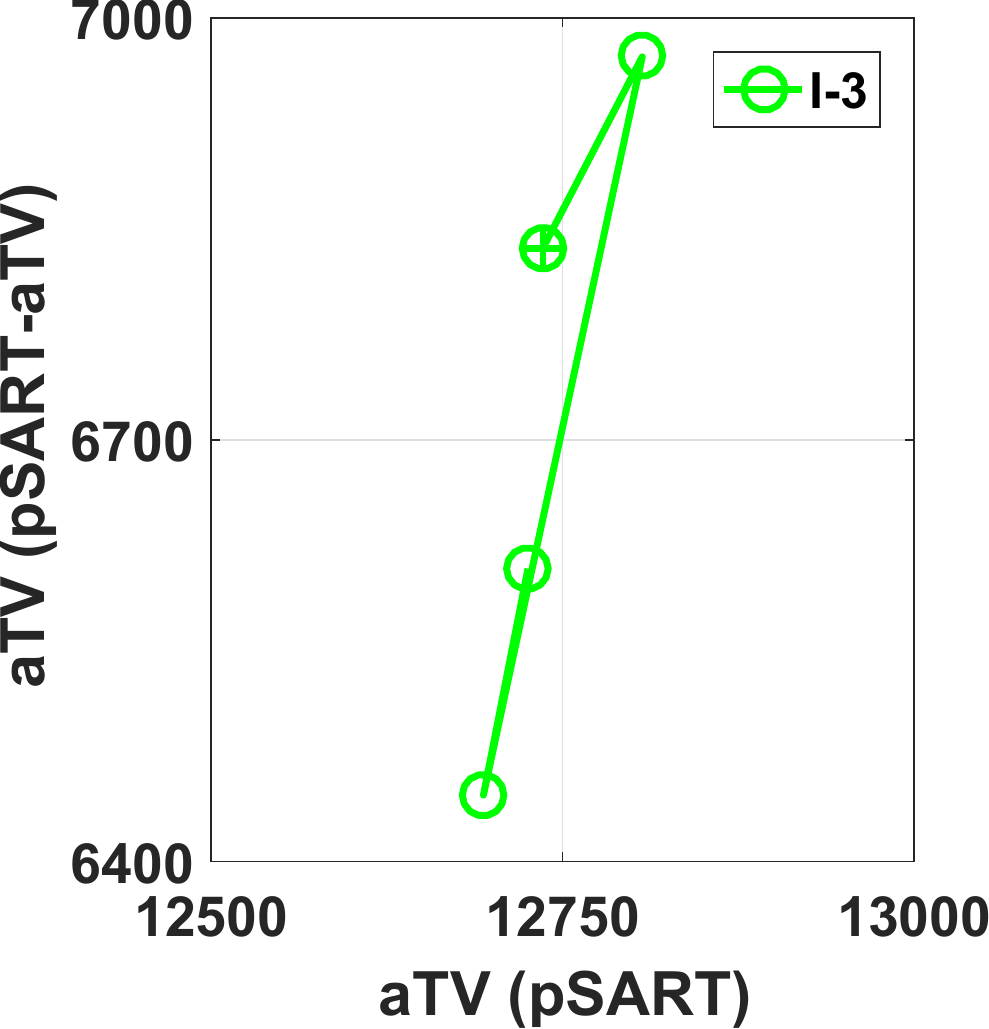}\\[10pt]
&\includegraphics[height=0.2\linewidth]{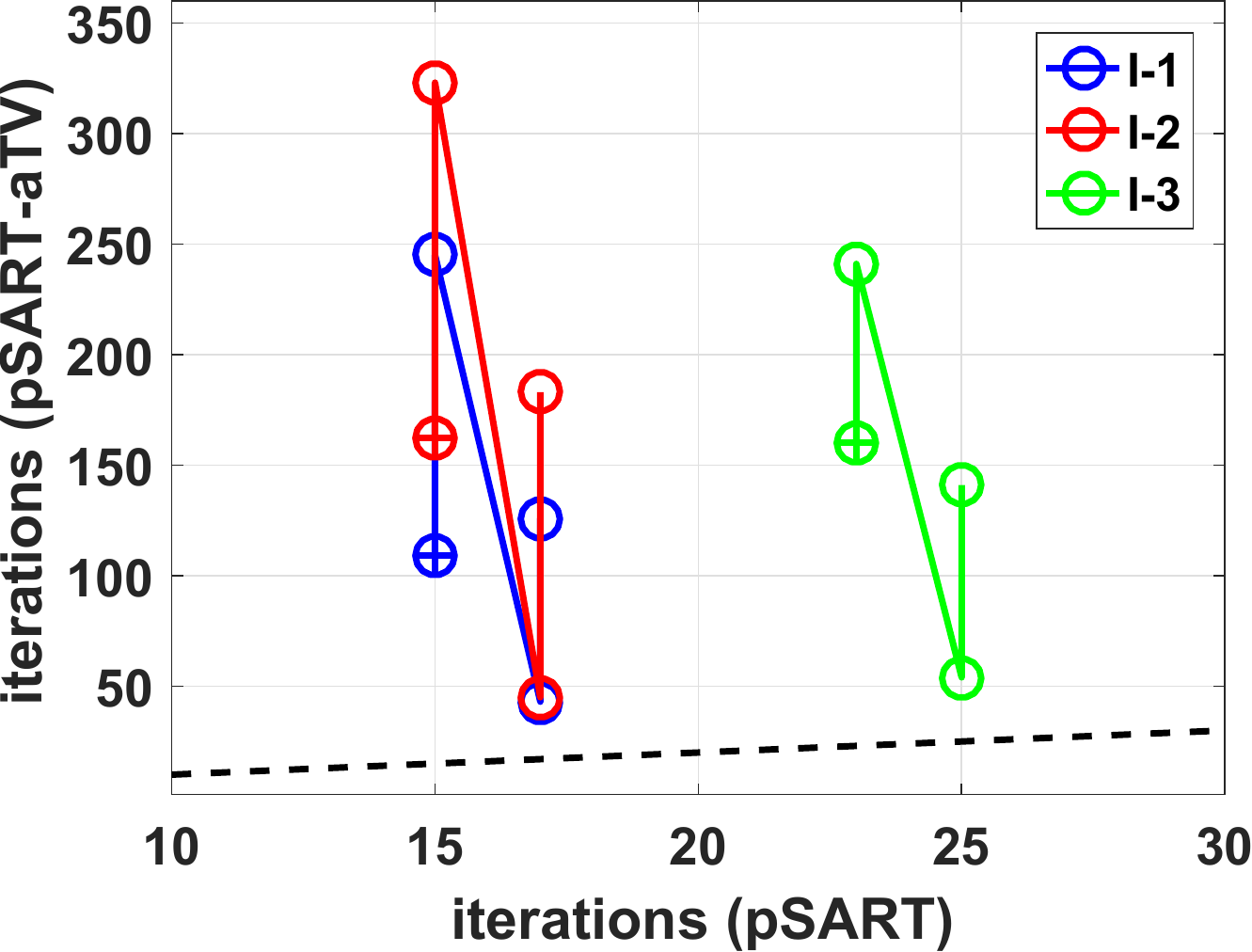} &\includegraphics[height=0.2\linewidth]{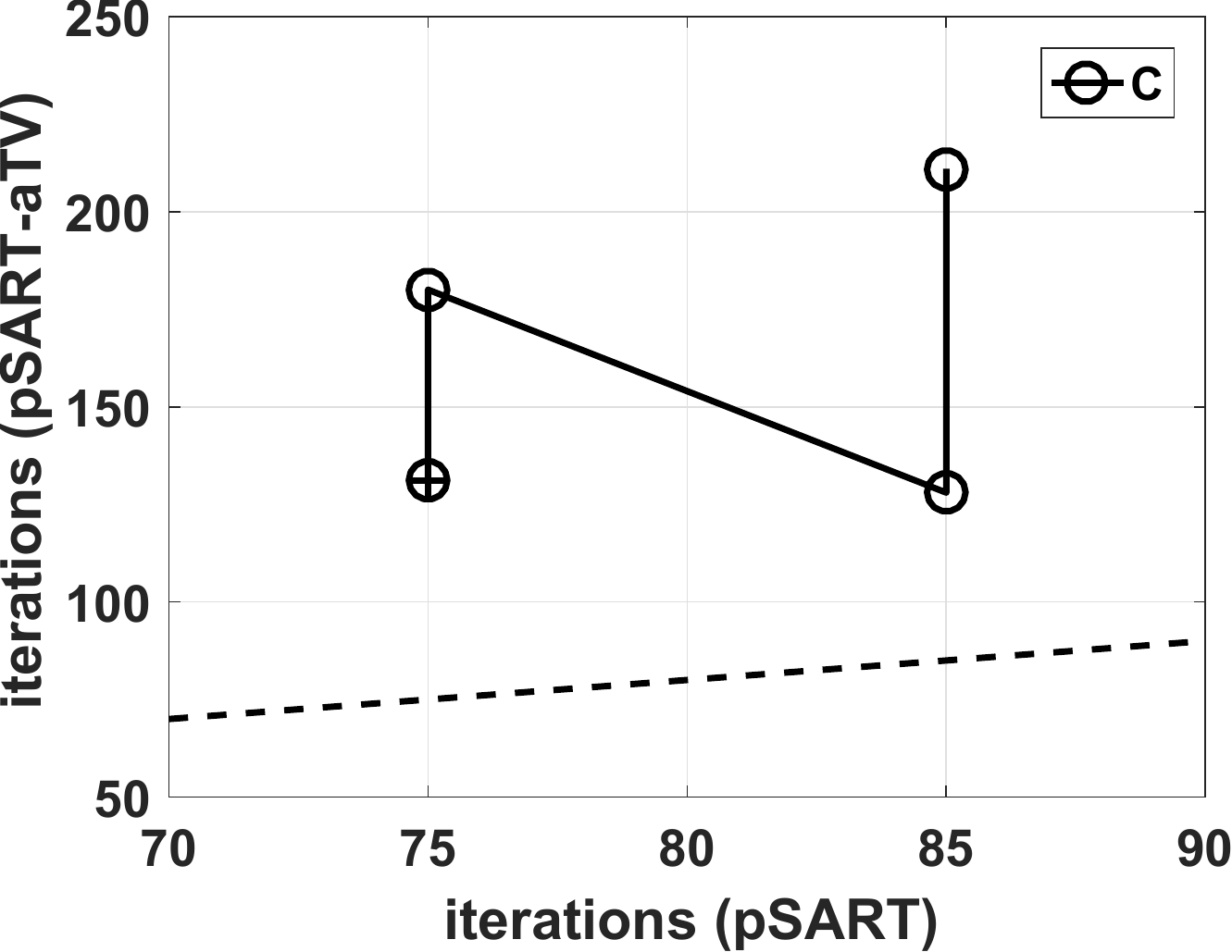}
\end{tabular}
\caption{Plots of important quantities from limited-angle experiments. Upper left plot: Values of $\varepsilon$ obtained by pSART and pSART-ATV for each simulation. Points marked with $+$ correspond to images in first and second columns of \Fref{F:limitedangleimages}, moving across that figure as one follows the line.  Upper middle plot: TV values for pSART reconstructions plotted against TV values for pSART-ATV reconstructions. Blue dashed line represents the average true ATV of the phantom at 70 keV.  Upper right plot: equivalent plot for data set I-3, shown on different axes. Bottom middle and left figures: Number of iterations required by each algorithm for each simulation.}\label{F:limitedangle_plots}
\end{figure}

\Fref{F:limitedangle_plots} shows a plot of relevant quantities from these experiments. As in the sparse-view simulations, pSART-ATV was able to find a solution with equal $\varepsilon$ value to pSART in every experiment, and with substantially lower ATV. In this case, the values of $\varepsilon$ obtained across different experiments (upper left plot) were fairly consistent, due in large part to the fact that the total number of measurements used to calculate the residual did not vary as much as in the sparse-view experiments. The plots of the ATV values (upper middle and right plots) indicate that the ATV values of the images reconstructed by pSART-ATV were substantially lower than those of the images reconstructed by pSART. For the three data sets using a 130 kVp spectrum, the ATV values of the reconstructed images were also fairly close to the average ATV of the true phantom at 70 keV (the ATV varies slightly across experiments as the weighting factors $\omega_i$ in \Eref{E:ATV1} depend on the angular extent and starting angle). For data set I-3, the average ATV of the true phantom was roughly 5730, somewhat lower than the reconstructed values.

The iterations required by pSART and pSART-ATV to obtain the same $\varepsilon$ value are shown in the bottom two plots. The ``N'' shape of the plot for all four data set indicates that the reconstructions from data acquired over the $0^\circ$ to $180^\circ$ arcs (fourth and eighth columns of \Fref{F:limitedangleimages}) generally required many more iterations of pSART-ATV to obtain an equivalent $\varepsilon$ value than the reconstructions from corresponding data acquired over the $90^\circ$ to $270^\circ$ arc (second and sixth columns). This is in spite of the fact that the $\varepsilon$ value in both cases was virtually the same. This phenomenon is difficult to explain, but is likely attributable to the differing nature of the artifacts that arose depending on which subset of angular data was missing. In general, the relationship between the value of $\varepsilon$ obtained by pSART and the number of iterations that pSART-ATV would require to obtain this value was difficult to predict. As desired, however, pSART-ATV was eventually able to obtain a $\varepsilon$-compatible solution in every case.

\subsection{Anatomical phantom}
As a more clinically realistic test case, we performed a second set of experiments using the XCAT anatomical phantom~\cite{SSMGT10}. The phantom size was $512 \times 512$ pixels with a pixel width of 0.75 mm. The simulation modeled a transaxial slice through the torso, including iodinated contrast agent in the bloodstream. The inclusion of iodine requires a modification to the pSART algorithm to determine whether to interpolate between soft tissue and iodine, or soft tissue and bone, within image pixels with LAC values higher than soft tissue~\cite{LS14b}. This is accomplished by a user-provided mask indicating which pixels are likely to contain bone, under the assumption that this can be determined {\em a priori}. 

Simulated data were generated using XCAT's analytic projection algorithm~\cite{SMBFT08}. The same 80 kVp spectrum was used as in the FORBILD phantom experiments, with noise proportional to $\int I_0(E) = 2 \times 10^5$. To model a realistic clinical acquisition, a fan-beam geometry was simulated with source to iso-centre distance of 80 cm, source-to-detector distance of 160 cm, and fan beam half-angle of 17.5$^\circ$. Since the fan-beam geometry is not symmetric over 180$^\circ$, we collected 900 views over a 360$^\circ$ rotation arc. The fan-beam system matrices used for the reconstruction algorithms were generated using the MIRT.

\begin{figure}
\begin{tabular}{ccc}
\includegraphics[width=0.25\linewidth]{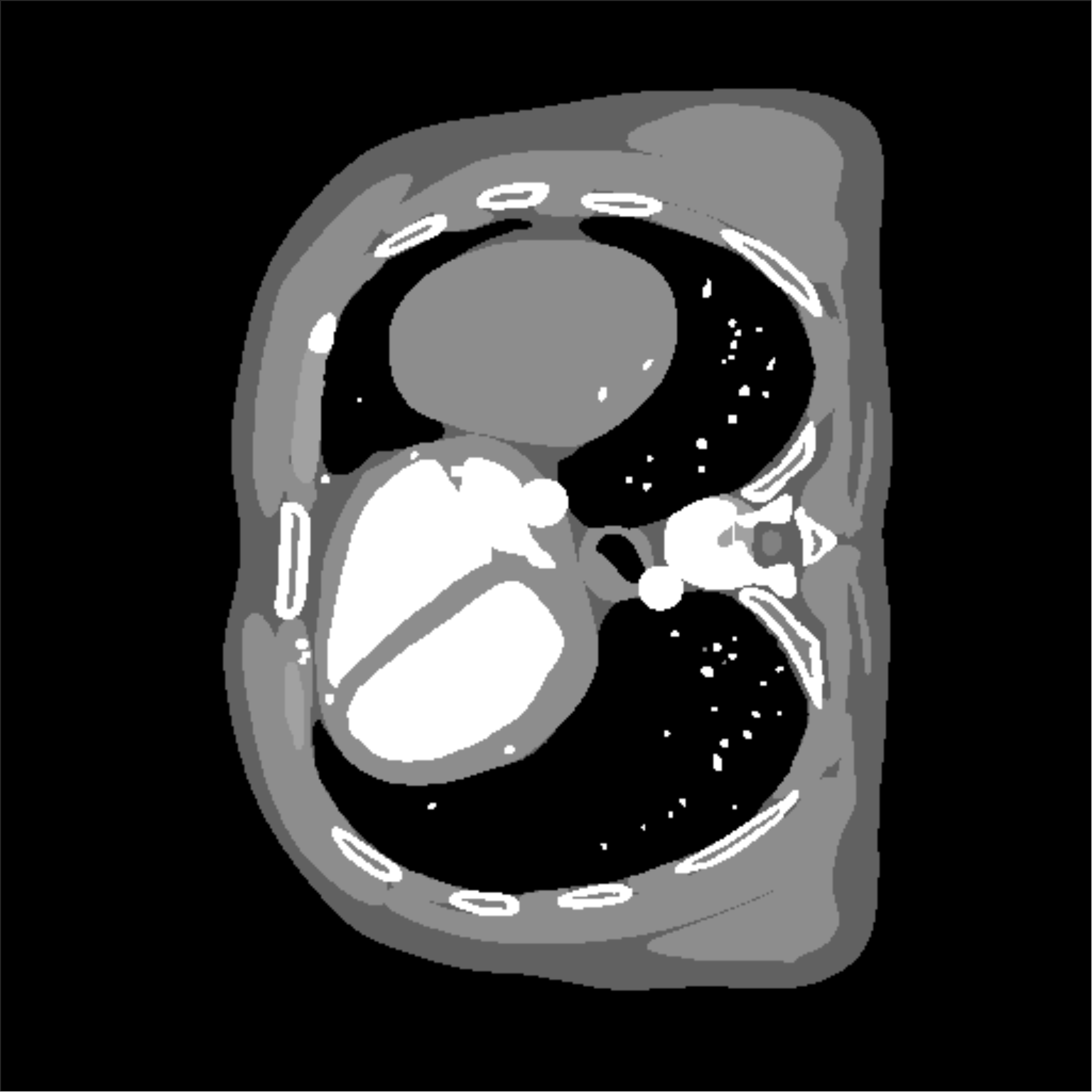} &\includegraphics[width=0.25\linewidth]{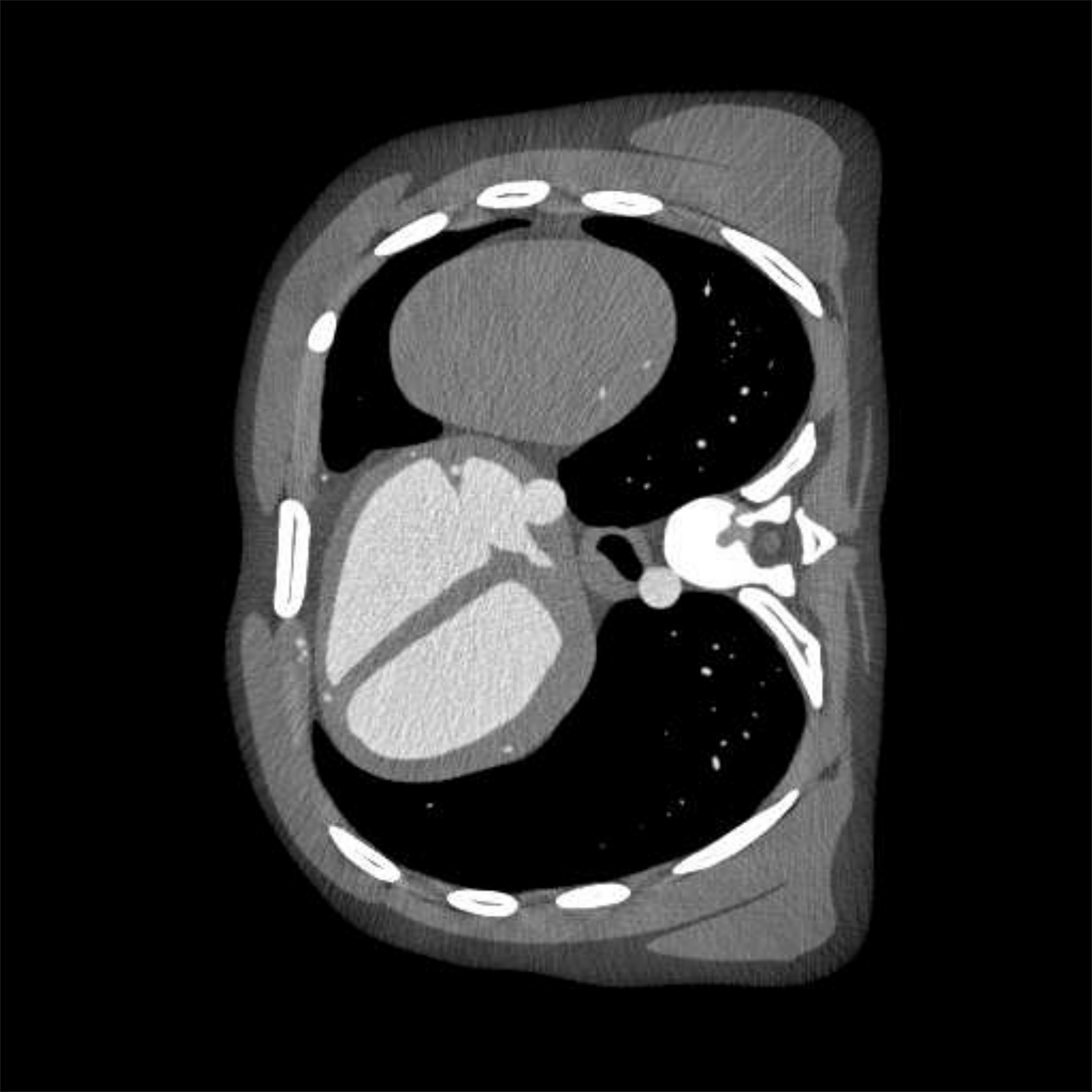} &\includegraphics[width=0.25\linewidth]{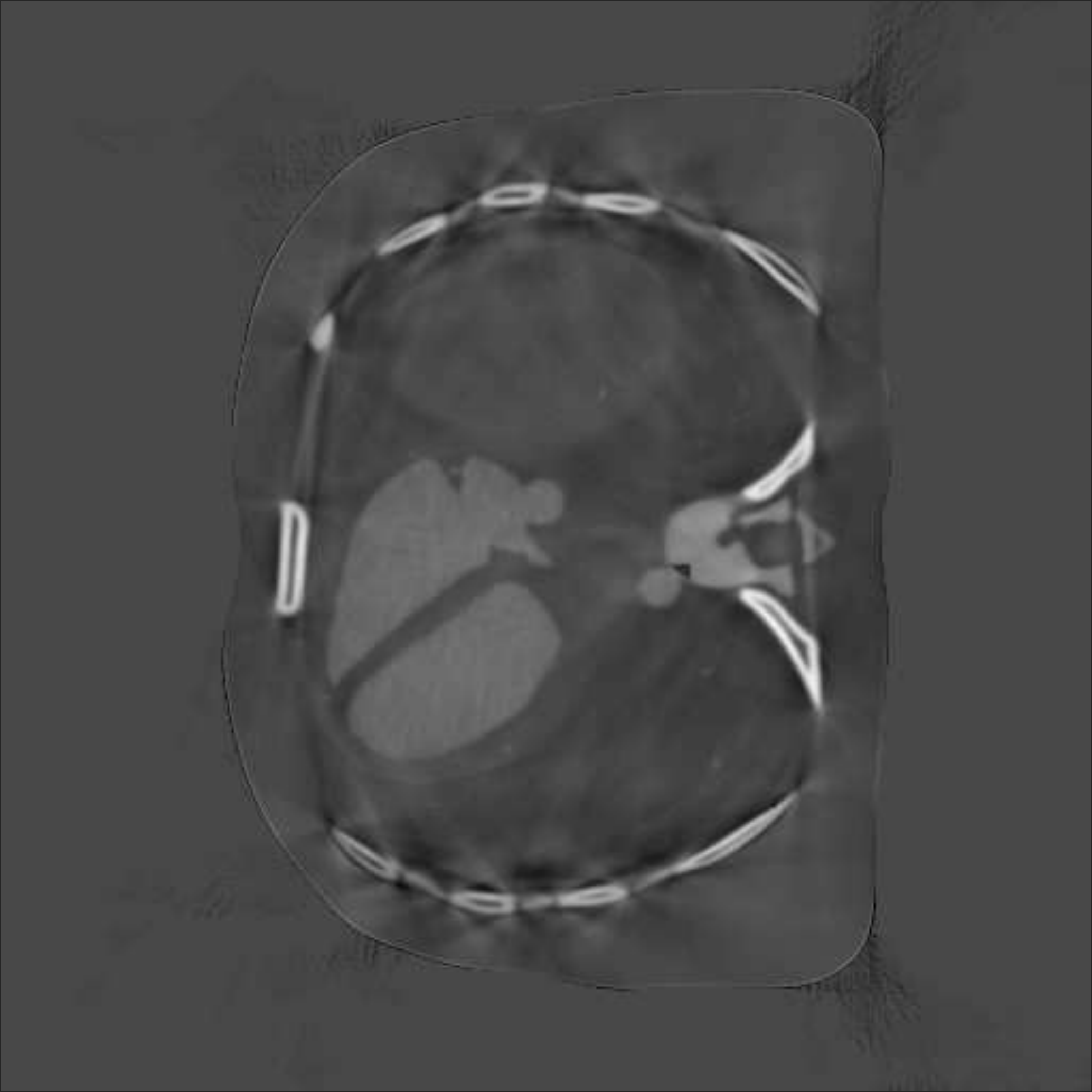}
\end{tabular}
\caption{XCAT anatomical phantom used for experiments with fan-beam data. Left: True phantom at reference energy of 50 keV. Phantom is displayed on a greyscale window of $[0.1, 0.35]$ cm$^{-1}$ to emphasize soft tissue contrast and artifacts in reconstructed images. Centre: image reconstructed using SART from water-corrected data. Right: Difference between image reconstructed using pSART (leftmost image in~\Fref{F:sparseview_images_XCAT}) and image reconstructed using SART. Greyscale window is $[-0.04, 0.10]$ cm$^{-1}$.}\label{F:xcat_phantom}
\end{figure}

An image of the phantom is shown in \Fref{F:xcat_phantom}, along with an image reconstructed by SART using soft tissue corrected data. While the beam hardening artifacts are not as apparent in this image as they were for the FORBILD phantom (due in part to the wider greyscale used to display the image), a subtraction of the soft tissue corrected SART reconstruction from the reconstruction obtained from the same dataset using pSART indicates that these artifacts are present in the former. In particular, there are streaking artifacts caused by bony structures such as the ribs and sternum, and the $\mu$ values of the body tissue and contrast agent tend to be underestimated.

For the sparse-view experiments, we reconstructed the phantom from 900, 225, 180, 90, and 45 views taken over $360^\circ$. As before, the number of iterations for which pSART was run scaled with the number of views, from 9 iterations for the 900-view dataset up to 180 iterations for the 45-view data set. The number of subsets was chosen such that each subset consisted of 15 views in all cases. pSART-TV was run with $\gamma = 0.9995$ and $N=60$.

\begin{figure}
\includegraphics[width=\linewidth]{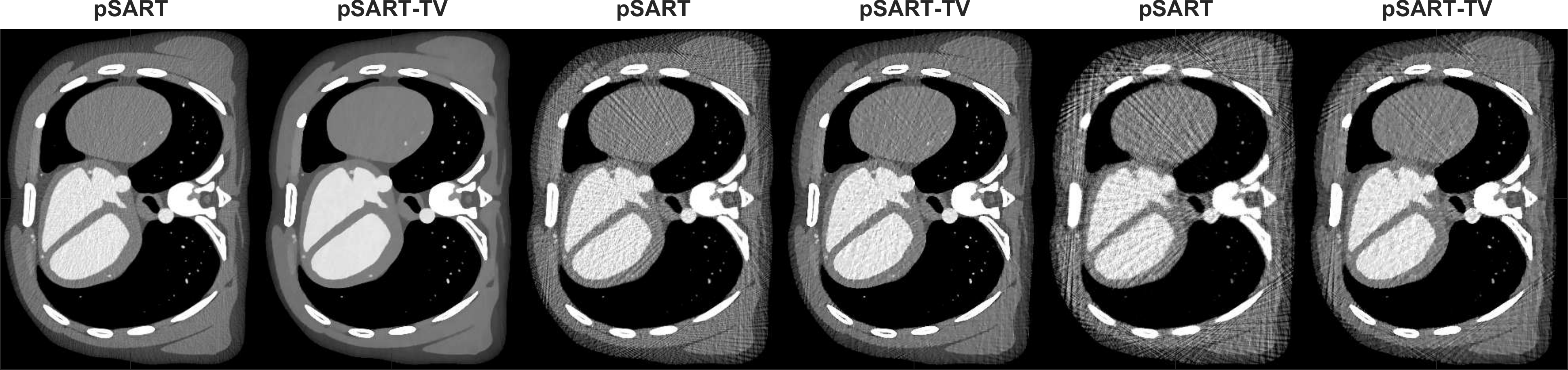}
\begin{tabular}{p{0.33\linewidth}p{0.33\linewidth}p{0.33\linewidth}} \centering $p = 900$ & \centering{ $p = 180$} & \centering{$p = 90$}
\end{tabular}
\caption{Images of XCAT phantom reconstructed from sparse-view data using pSART and pSART-TV. First, third and fifth images are reconstructed with pSART, while second, fourth and sixth images are corresponding reconstructions with pSART-TV.}\label{F:sparseview_images_XCAT}
\end{figure}

\begin{figure}
\begin{tabular}{ccc}
\includegraphics[height=0.2\linewidth]{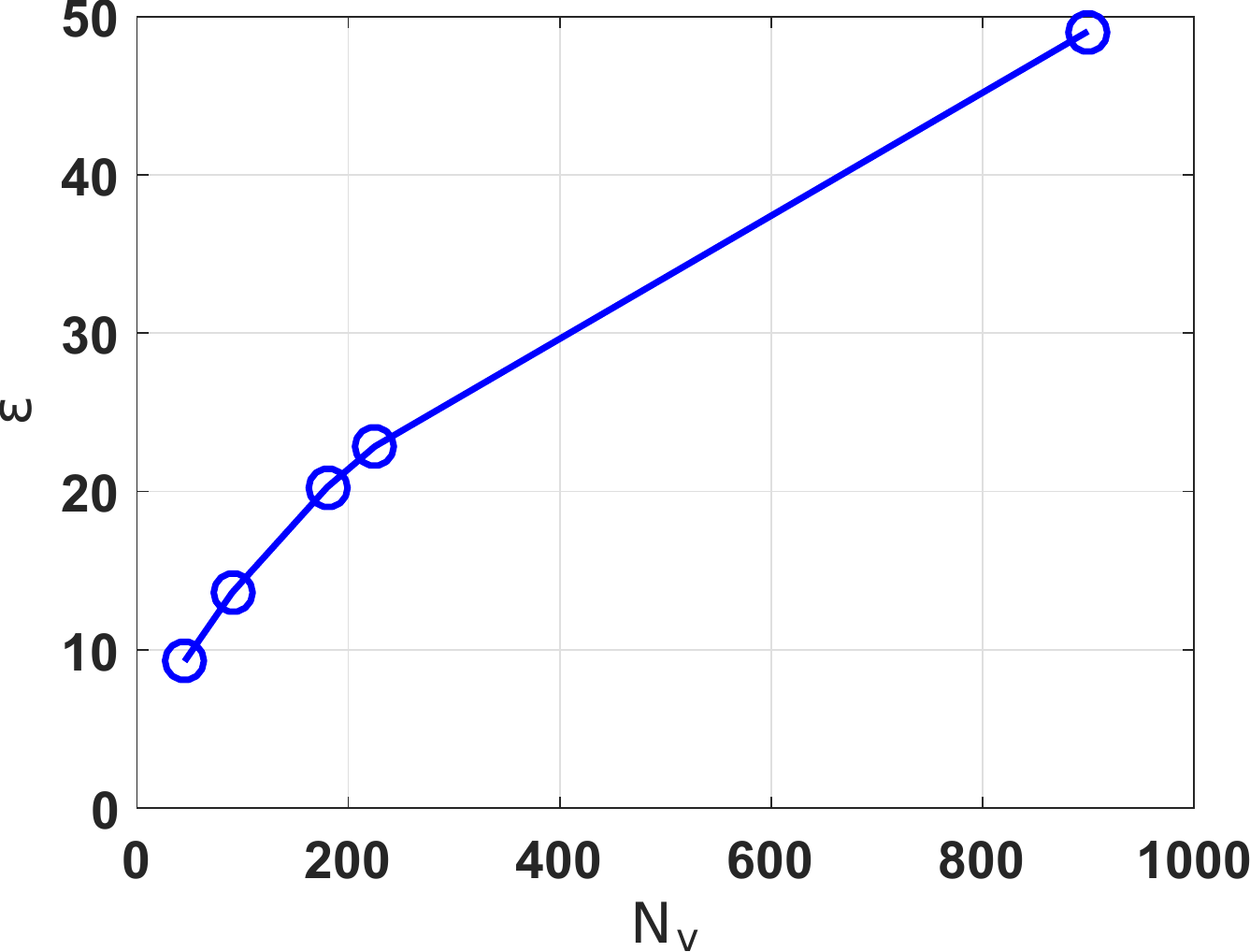} &\includegraphics[height=0.2\linewidth]{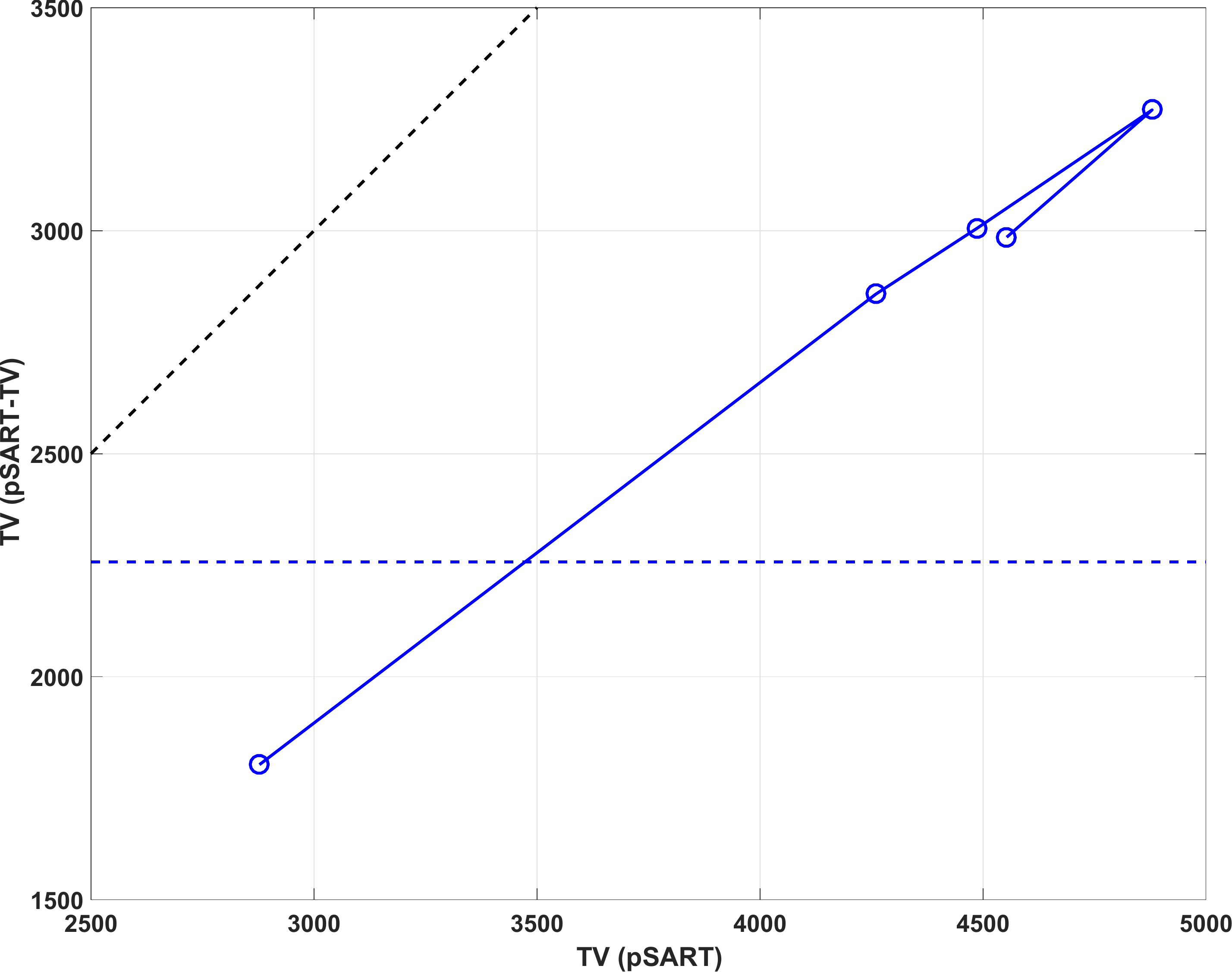} &\includegraphics[height=0.2\linewidth]{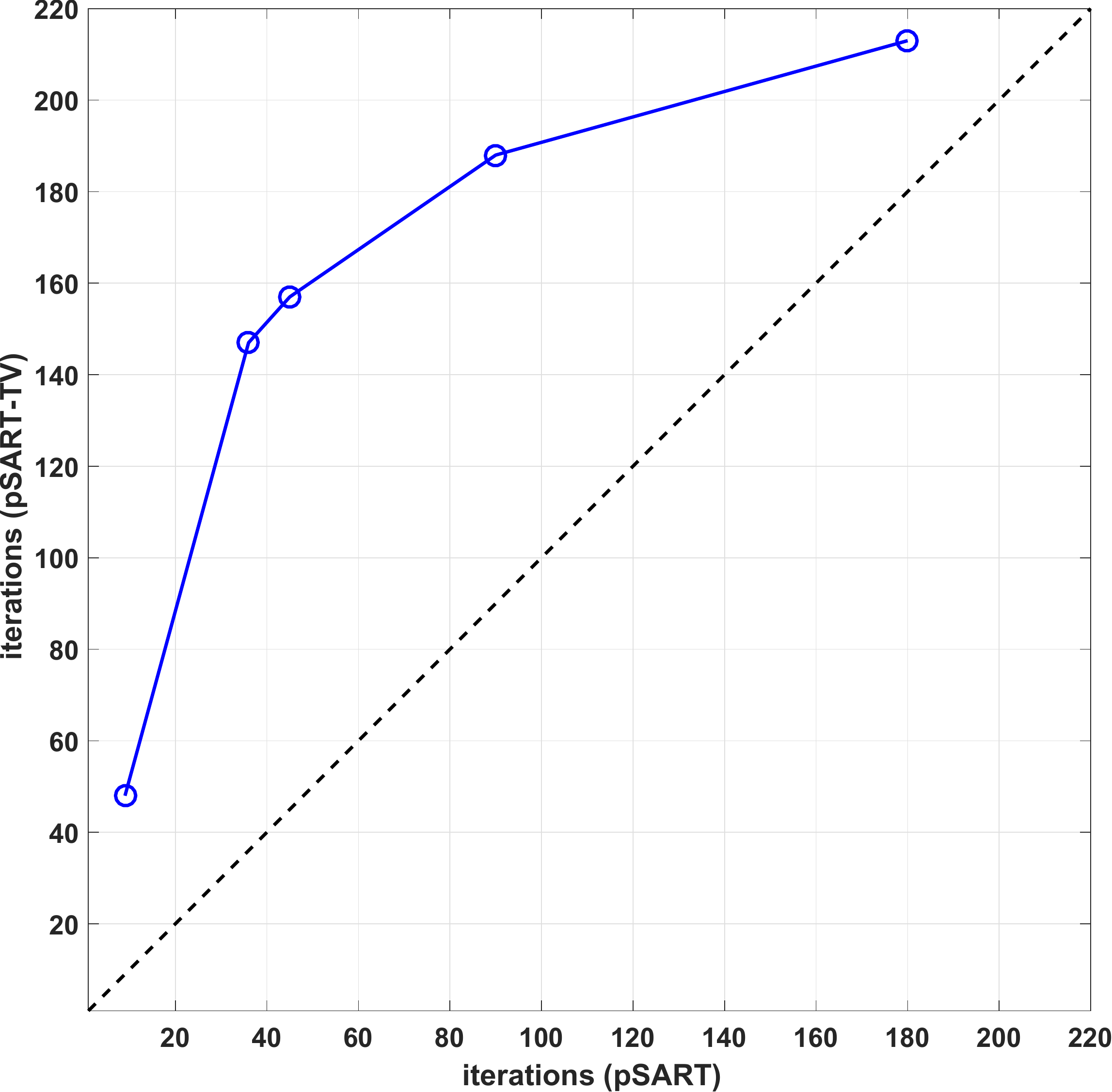}
\end{tabular}
\caption{Plots of important quantities from sparse-view experiments with XCAT phantom. Left plot: Values of $\varepsilon$ obtained by pSART and pSART-TV. Middle plot: TV values for pSART reconstructions plotted against TV values for pSART-TV reconstructions. Black dashed line is the line $y=x$, blue dashed line represents true TV of phantom at 50 keV.  Leftmost point corresponds to $p=900$. Right plot: number of iterations required by each algorithm for each simulation. Leftmost points correspond to $p=900$.}\label{F:sparseview_plots_XCAT}
\end{figure}

Some representative images are shown in~\Fref{F:sparseview_images_XCAT}. With 900 views, the effect of pSART-TV is primarily to reduce some of the noise in the image. As the number of views is reduced, undersampling artifacts become more prevalent in the reconstructions generated using pSART. The superiorized algorithm is successful in reducing these artifacts, although they are not entirely removed. \Fref{F:sparseview_plots_XCAT} shows plots of the $\varepsilon$ values, TV values and number of iterations required for each of the reconstructions. The trends are consistent with the experiments conducted with the FORBILD phantom; pSART-TV found a solution that was as constraints-compatible as the image reconstructed by pSART, with significantly lower TV values. As before, obtaining this solution required running many more iterations of the algorithm, in general.

For the limited-angle scenarios, we were not able to use pSART-ATV in this experiment, since the weights used in the computation of the ATV~\eref{E:ATV1} are calculated based on a parallel-beam geometry~\cite{CJLW13}. Instead, we used pSART-TV; while~\cite{CJLW13} indicates that an ATV penalty is more effective in removing limited angle artifacts, TV is also able to reduce these artifacts considerably. As in the previous experiments, we simulated two acquisitions over 165$^\circ$ and two over 150$^\circ$, with different starting points. Other parameters to the reconstruction algorithms were the same as in the sparse-view experiments.

\begin{figure}
\includegraphics[width=\linewidth]{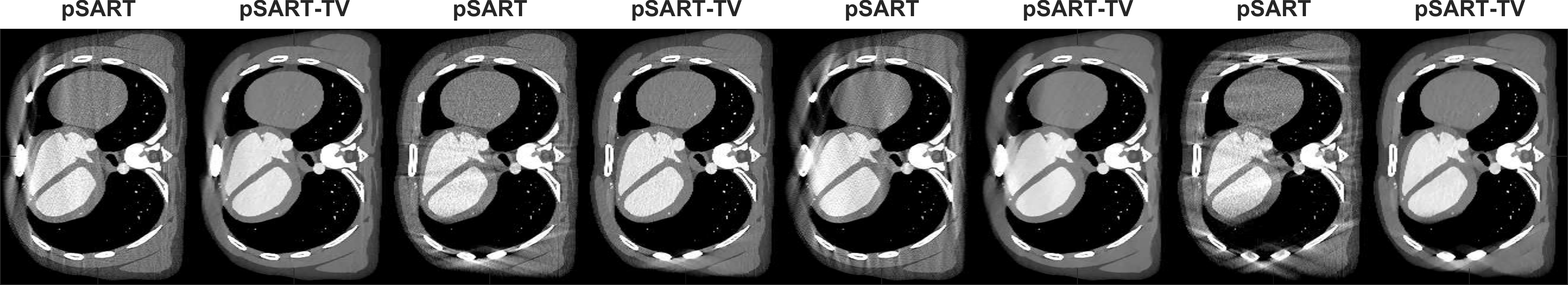}
	\begin{tabular}{p{0.25\linewidth}p{0.25\linewidth}p{0.25\linewidth}p{0.25\linewidth}}
		Extent: 165$^\circ$ & Extent: 165$^\circ$ & Extent: 150$^\circ$ & Extent: 150$^\circ$\\
		Start: $97.5^\circ$	& Start: $7.5^\circ$ & Start: $105.0^\circ$ & Start: $15.0^\circ$ 	
	\end{tabular}
\caption{Images of XCAT phantom reconstructed from limited angle data using pSART and pSART-TV.  First, third, fifth and seventh images are reconstructed with pSART; second, fourth, sixth and eighth images are corresponding images reconstructed with pSART-TV.}\label{F:limitedangle_images_XCAT}
\end{figure}

\begin{figure}
\begin{tabular}{ccc}
\includegraphics[height=0.2\linewidth]{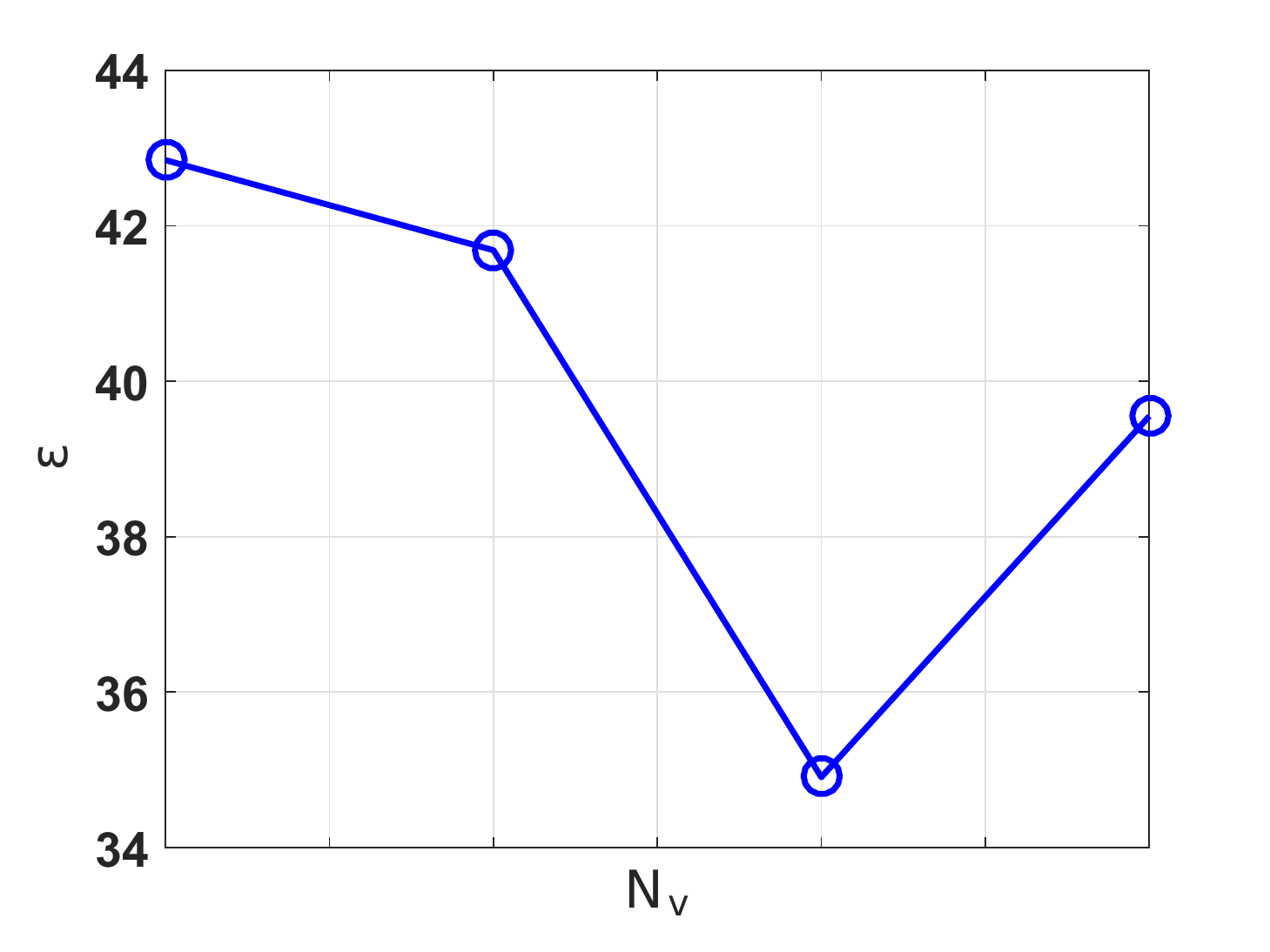} &\includegraphics[height=0.2\linewidth]{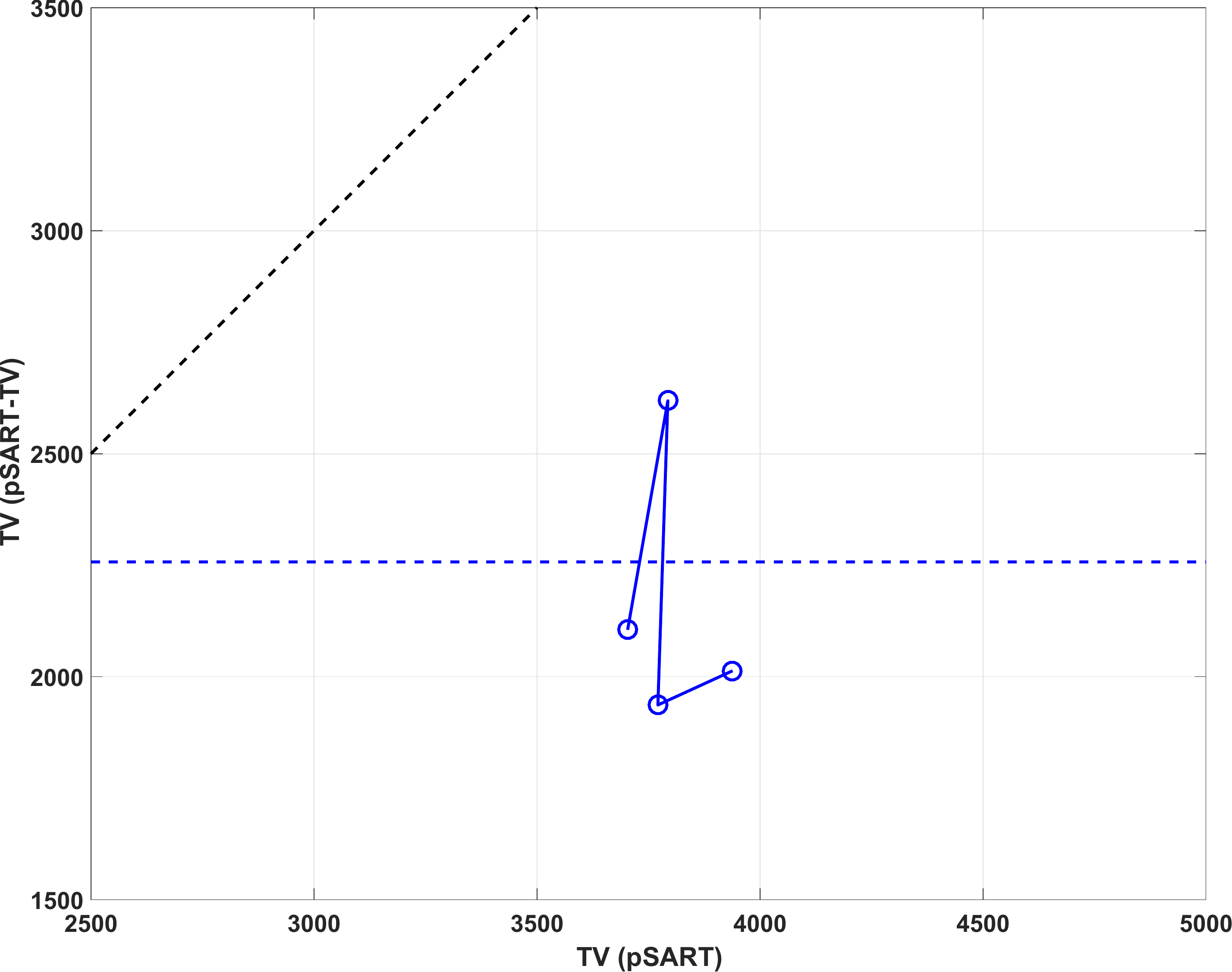} &\includegraphics[height=0.2\linewidth]{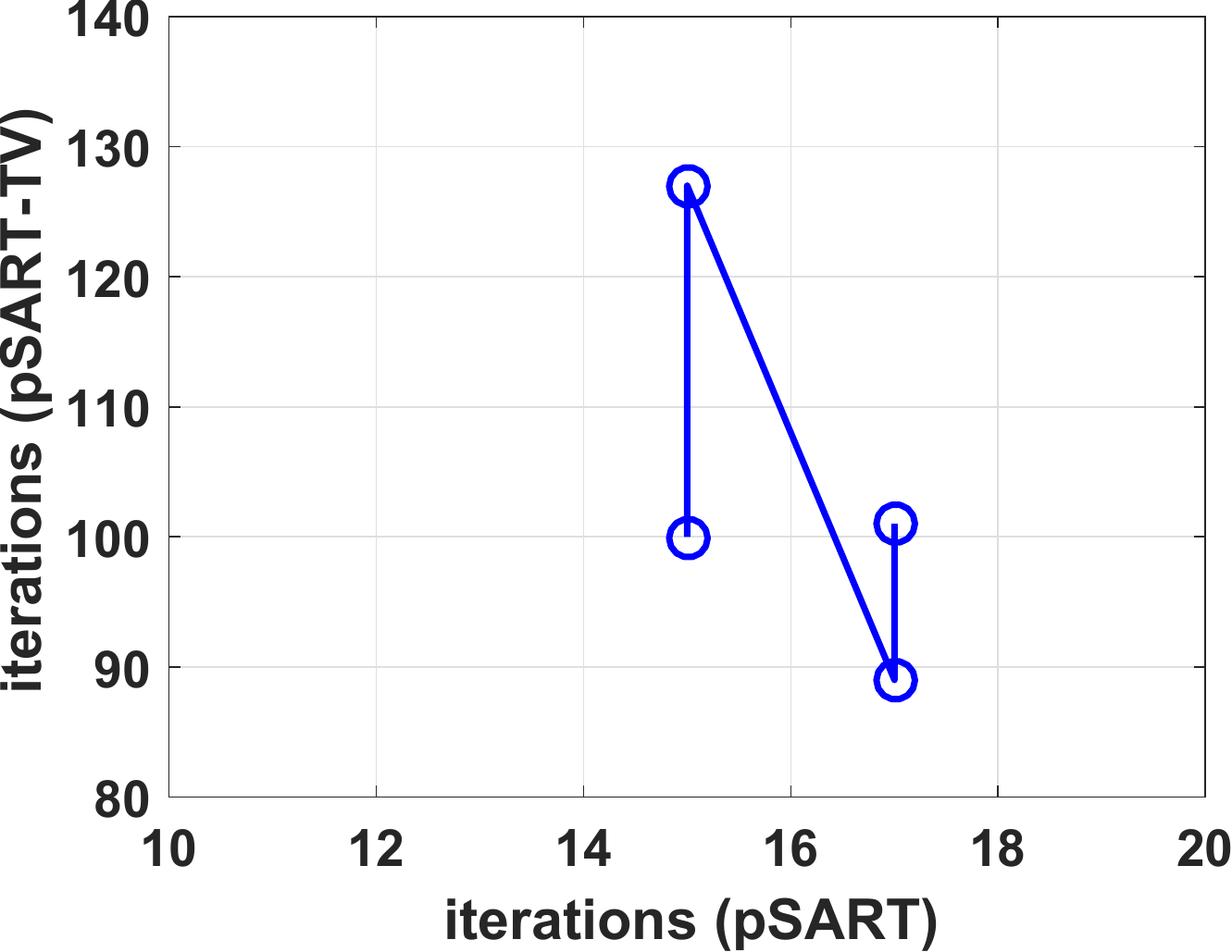}
\end{tabular}
\caption{Plots of important quantities from limited angle experiments with XCAT phantom. Plotted quantities are the same as for \Fref{F:sparseview_plots_XCAT}. Leftmost point corresponds to first and second columns of ~\Fref{F:limitedangle_images_XCAT} in all cases, moving across that figure as one follows the line.}\label{F:limitedangle_plots_XCAT}
\end{figure}

Representative images and plots of important quantities are shown in Figures~\ref{F:limitedangle_images_XCAT} and \ref{F:limitedangle_plots_XCAT}, respectively. It is apparent from the reconstructed images that pSART-TV is effective in reducing the limited angle artifacts that are present in the corresponding images reconstructed by pSART. In the two experiments with 165$^\circ$ angular extents, the artifacts are largely eliminated aside from a few in the vicinity of the sternum and ribs. When the angular extent is further reduced to 150$^\circ$, the severity of the artifacts increases, and pSART-TV is less effective in removing them. As the plots in Figure~\ref{F:limitedangle_images_XCAT} indicate, however, the superiorized algorithm is again able to find a $\varepsilon$-compatible solution with substantially lower TV in every case.

\section{Conclusions}\label{S:conclusions}
In this paper we present a superiorized algorithm for reconstruction of CT images from sparse-view and limited-angle polyenergetic data. The primary contribution of the paper is the application of the superiorization methodology to an iterative algorithm for reconstruction of images from polyenergetic data, whereas previously it has been applied to algorithms which assume a monoenergetic model. The polyenergetic reconstruction algorithm is pSART~\cite{LS14b}, and the chosen objective functions for superiorization are total variation (TV) and anisotropic TV (ATV)~\cite{CJLW13}, which have been successfully applied in the past to reconstruction of sparse-view and limited-angle monoenergetic data. Our superiorized algorithm significantly reduces artifacts that arise from sparse-view and limited-angle data, as well as the beam hardening artifacts that would be present if a superiorized version of a monoenergetic algorithm were used.

A sufficient condition to guarantee success of a superiorized algorithm is for the algorithm to be strongly perturbation resilient. We are not able to establish that pSART is a strongly perturbation resilient algorithm, as the conditions under which the algorithm converges are not well understood~\cite{H15}. In our numerical experiments, however, the superiorized pSART algorithm is successful in every instance. It produces solutions that are as constraints-compatible as those produced by pSART, with TV or ATV values that are typically 30 -- 60\% lower. As the authors note in~\cite{HGDC12}, strong perturbation resilience is a sufficient condition for the success of superiorization, but not a necessary one. Thus, the inability to prove whether or not an algorithm is strongly perturbation resilient does not preclude the use of superiorization to obtain solutions that are improved with respect to the chosen objective.

Whether the image is superior with respect to the chosen objective is, of course, a separate question from whether the image quality is improved; this depends on how effective the objective function is at penalizing undesirable image characteristics. In our experiments we find that TV and ATV are effective at eliminating artifacts caused by sparse-view and limited-angle data, up to a point. For the FORBILD phantom, we were able to obtain reconstructions that were largely free of artifacts from 165$^\circ$ of data in limited angle experiments, and roughly one quarter to one fifth the angular sampling rate that (conservatively) would be required to reconstruct the phantom at a resolution of $800 \times 800$ pixels, in our sparse-view experiments. Results of a second experiment using the XCAT anatomical phantom with a simulated fan-beam acquisition gave similar results, showing that the approach is not dependent on a parallel-beam geometry. %It is reasonable to expect that these limitations are object-specific; an accurate reconstruction from less data might be possible for a simpler object, while an object with more complex geometry would require more. 

While we were successful in superiorizing the pSART algorithm, the pSART-TV and pSART-ATV algorithms are not without drawbacks. A primary deficiency is the increased computation time required to find an $\varepsilon$-compatible solution. In addition to the extra cost per iteration to reduce the objective function value, it may take many more iterations than the original algorithm to find such a solution. This was particularly true in the limited-angle experiments, where it often required more than twice as many iterations to find an $\varepsilon$-compatible solution (with consistent data) and sometimes more than ten times as many iterations (with inconsistent data). This may be due to the choice of the parameters $\gamma$ and $N$ in these experiments, which we found needed to be larger than in the sparse-view experiments to effectively remove artifacts. Additionally, because the pSART algorithm is an algebraic technique, it does not model noise as accurately as a statistical technique. Superiorizing a statistical polyenergetic algorithm (e.g.~\cite{DNDMS01}) and using a likelihood-based proximity function, as was done in~\cite{GH14}, could provide better results in the case of noisy data.

\section*{References}

\bibliographystyle{dcu}

\end{document}